\documentclass[useAMS,usenatbib]{./TexStyles/mn2e}
\usepackage{natbib}
\usepackage{graphicx}
\usepackage{floatflt}
\usepackage{amssymb}
\newcommand{\comment}[1]{}
\title[Optical emission line properties of Narrow Line Seyfert 1s and comparison AGN]{Optical emission line properties of Narrow Line Seyfert 1s and comparison AGN}
\author[J. R. Mullaney and M. J. Ward]{J. R. Mullaney$^{1}$\thanks{E-mail:
j.r.mullaney@dur.ac.uk} and M. J. Ward$^{1}$\\
$^{1}$Institute for Computational Cosmology, Durham University, South Road, Durham, DH1 3LE, U.K.}
\begin{document}

\date{Date Accepted}

\pagerange{\pageref{firstpage}--\pageref{lastpage}} \pubyear{2007}

\maketitle

\label{firstpage}

\begin{abstract}
Based on a new spectroscopic sample observed using the WHT, we examine the kinematic properties of the various emission line regions in narrow line Seyfert 1 galaxies (NLS1s) by modelling their profiles using multiple component fits.  We interpret these results by comparison with velocity components observed for different lines species covered in the same spectrum, and equivalent components measured in the spectra of some broad line Seyfert 1s and a representative Seyfert 2 galaxy.  We find that the fits to the H$\alpha$ and H$\beta$ line profiles in NLS1s require an additional broad ($\gtrsim$3000km/s) component that might correspond to a suppressed broad line region with similar kinematics to those of typical broad line Seyfert 1s.  From the profiles of the forbidden high ionisation lines (FHILs) in NLS1s, we find evidence that they appear to trace an `intermediate' velocity region with kinematics between the standard broad and narrow line regions.  Weaker evidence of this region is also present in the profiles of the permitted Balmer lines.  Finally, we note that despite having similar ionisation potentials, the relative intensities of the highly ionised lines of [Fe X]$\lambda$6374 and [FeXI]$\lambda$7892 show considerable dispersion from one galaxy to another. The interpretation of this requires further modelling, but suggests the possibility of using the ratio as a diagnostic to constrain the physical conditions of the FHIL emitting region and possibly the shape of the spectral energy distribution in the vicinity of 200eV.  This spectral region is very difficult to observe directly due to photoelectric absorption both in our Galaxy and intrinsic to the source.

\end{abstract}


\section{Introduction}
Narrow line Seyfert 1 (NLS1) galaxies are a sub-set of the AGN population which share a number of characteristic properties \citep{Pogge00}. Superficially, their optical spectra resemble those of type 2 Seyferts in that their permitted emission lines are quite narrow and only marginally wider than their forbidden lines - a property  that has unfortunately led to numerous misidentifications as Sy2s in the literature.    However, \cite{Osterbrock85} noted that their strong permitted Fe II lines and high H$\beta$/[O III]$\lambda$5007 ratios were similar to the traditional type 1 Seyferts.  Although the velocity widths of the permitted lines in AGN spectra are observed to cover a continuous range ($\sim$10$^{2}$ to $\sim$10$^{4}$ km/s), \cite{Goodrich89} placed a somewhat arbitrary limit of to the FWHM of H$\beta$ of 2000km/s, for an object to be classed as a NLS1. Since then most studies have continued to use this value as the boundary between NLS1s and their broad line counterparts, here referred to as broad line  Seyferts, or BLS1s, in order to distinguish them from NLS1s.

Studies of the continuum emission from NLS1 revealed interesting features across the electromagnetic spectrum, for example, significant soft X-ray excesses, rapid X-ray variability and strong emission from high ionisation forbidden lines (FHILs) (e.g. \citealt{Stephens89}; \citealt{Puchnarewicz92}; \citealt{Pogge00}). Their X-ray properties encouraged searches using X-ray surveys with follow-up confirmation from optical spectroscopy. This showed that around 50\% of all \textit{ROSAT} soft X-ray selected AGN were NLS1s (\citealt{Grupe94}; \citealt{Pogge00}).  It should be noted, however, that this method introduces serious selection effects, since \cite{Boller96} found that not all NLS1s display strong soft X-ray excesses.

The features described above, in particular the rapid X-ray variability and lack of significantly broadened permitted emission lines has led to speculation that NLS1s may harbour a lower mass central black hole than is typical in BLS1s (10$^{7-8}$M$_{\odot}$ and 10$^{8-9}$M$_{\odot}$, respectively; e.g. \citealt{Mathur01}).  This, coupled with their bolometric luminosities,  implies a high accretion rate, possibly super-Eddington suggesting that the key parameter in determining the distinctive physical properties of this class is their high ratio of $\dot{m}/m$.  Indeed high luminosity AGN which do not fulfil the velocity width criteria of NLS1s (i.e. $<$2000km/s) but share many of their other characteristics are also thought to have high $\dot{m}/m$, e.g. PDS456 (\citealt{Reeves03}), consistent with the model of \cite{McHardy06}.  It is partly for this reason that NLS1s have attracted much interest. In addition, we can potentially test the full extent of the relationships between observable parameters that have been established for the more general AGN population, such as relations between black hole mass and stellar bulge luminosity and velocity dispersion (see e.g. \citealt{Ferrarese00}; \citealt{Gebhardt00b}).  Also, evidence of some of the more dramatic features of AGN activity, in particular relativistic and non-relativistic outflows, are sometimes  prominent in NLS1 spectra (e.g. \citealt{Leighly01}; \citealt{Leighly04}), although it is unclear whether this is an indication of a real propensity toward outflows in NLS1s, or whether other spectral features simply allow these features to be more easily measured in these AGN.  This has encouraged their study as laboratories for some phenomena present in general in AGN, but exhibited in a more extreme form in the NLS1s. 

The optical spectra of NLS1 provide a wealth of information. The large number of permitted and forbidden lines give  information on the kinematics and physical properties of the various emitting regions, as well as the shape of the (unobscured) ionising continuum.  For example, \cite{Dietrich05} compared the low ionisation ($<$100eV) emission lines of NLS1 to show that the far-UV continuum of NLS1 was not significantly different from that of BLS1s, thereby indicating that the soft X-ray excess, seen in many NLS1s, is unlikely to extend to significantly lower energies approaching 100eV.  However the true extent of this continuum feature, which is likely to contribute a significant fraction of the bolometric luminosity in a large proportion of NLS1s, is still unknown.

 With ionisation potentials greater than 100eV, FHILs offer the possibility of determining the true shape of the soft X-ray excess. Previous studies of the optical FHILs in NLS1s, most notably [Fe VII]$\lambda$6087 (ionisation energy: 99eV), [Fe X]$\lambda$6374 (235eV), [Fe XI]$\lambda$7892 (262eV) and [Fe XIV]$\lambda$5303 (361eV) noted that these lines were significantly blueshifted and broadened, with respect to the low-ionisation forbidden lines, suggesting the possibility of outflows between the traditional broad- and narrow- line regions (e.g. \citealt{Erkens97}; \citealt{Porquet99}; \citealt{Nagao00}).  However, despite the fact they  often display broad wings and have prominent profile asymmetries, little effort has been directed towards modelling the FHILs using multiple component fits. This could yield insights into the detailed kinematics of the highly ionised gas close to the AGN.  The information gained from gas kinematics would help us to constrain the location of the FHIL emitting gas with respect to other emitting regions, and also help us obtain a better  understanding of outflows from AGN.

In this paper, we examine the optical emission line spectra of 10 nearby (z $<$ 0.1) AGN at sufficiently high spectral resolution and signal to noise to permit us to model their FHILs using multiple components.  Six of these AGN have previously been classified as NLS1s.  We also observed three BLS1s to facilitate comparison between the Seyfert groups.  In an attempt to determine whether the FHILs are affected by the obscuration of the putative dusty molecular torus, we also observed one Seyfert 2 galaxy, Mrk573, which has been shown to contain a hidden broad line region by \cite{Nagao04}.  The high quality of our spectra also enable accurate multiple component fits to both the forbidden and permitted emission lines, allowing precise measurement of the kinematics of the broad and narrow emission line regions in NLS1s and BLS1s.  This paper is structured as follows: in \S2 we discuss the observations and spectral reduction techniques.  We present our main results, including measurement of the FHILs, in \S3 and in \S4 we interpret our findings in terms of current AGN models. We conclude with a summary our principal results in the final section.

\section{Observations}

The ten AGN in our sample were observed using the ISIS long slit spectrograph on the William Herschel Telescope during the nights of the 17$^{th}$ and 18$^{th}$ of October 2006.  Basic information on these objects is listed in Table \ref{ObjTable}.  Both the red and blue arms of the ISIS instrument were used, with a 600 lines/mm grism in each case.  The CCD detectors used were the EEV12 and the Marconi CCD in the blue and red arms respectively.  In order to cover the wavelengths of all four principal optical FHILs ([Fe VII], [Fe X], [Fe XI], [Fe XIV]) using this instrument, it was necessary to observe each galaxy with two different grating angles in one of the arms.  Based on previous studies of FHILs in NLS1 galaxies (\citealt{Erkens97}), it was apparent that the [Fe XIV]$\lambda$5303 line was likely to be the weakest species of those observed. For this reason this line was observed twice using the blue arm, whilst the grating angle in the red arm was initially set for the  measurement of the [Fe VII]$\lambda$6087 and [Fe X]$\lambda$6374 species, followed by a grating angle change to allow measurement of the [Fe XI]$\lambda$7892 line at a longer wavelength.  This procedure thus resulted in a longer integration time for the weaker [Fe XIV] line than for the other iron species. The choice of beam-splitting dichroic was based on the observed wavelengths of the FHILs.  Objects with $z\leq0.035$ were observed using the 5700\AA \ dichroic, while objects of $z>0.035$ were observed using the 6100\AA \ dichroic.  This choice ensured that the lines of most interest always lay in a wavelength  region of high dichroic throughput.  The instrumental set-up provided almost complete spectral coverage between 4800\AA \ and 8900\AA.  In order to minimise the stellar contribution from the host galaxy we used a slit width of $0.5\arcsec$, resulting in an instrumental spectral resolution of $R=\lambda/\Delta \lambda =c/\Delta v \approx 3500-5500$, depending on wavelength (equivalent to 85-55 km/s FWHM). Corresponding standard star and telluric dividing star frames were taken for each galaxy observation.  Wavelength calibration frames were  taken using a CuNe + CuAr arc lamp.
 
\begin{table*}
\centering
\begin{tabular}{lcccccccc}
\hline
\hline
Galaxy&Type&z&R.A.&Dec&\multicolumn{3}{c}{Exposure Times (s)}&Ref.\\
 & & &\multicolumn{2}{c}{(J2000)} & Blue&Red/Short&Red/Long&\\
\hline
Ark 564&NLS1&0.0247&22:42:39.3&+29:43:31&1080&540&540&1\\
1H1934-063&NLS1&0.0106&19:37:33.0&-06:13:05&1800&900&900&1\\
IIZW136&NLS1&0.0633&21:32:27.8&+10:08:19&2400&1200&1200&2,3\\
IZW1&NLS1&0.0611&00:53:34.9&+12:41:36&2400&900&1200&4\\
Mrk335&NLS1&0.0261&00:06:19.5&+20:12:10&1600&600&1000&1,6\\
Mrk573&Sy2&0.0172&01:43:57.8&+02:21:00&1800&900&900&5\\
Mrk618&NLS1&0.0354&04:36:22.2&-10:22:34&3000&1500&1500&6\\
NGC985&Sy1&0.0432&02:34:37.8&-08:47:15&1800&1200&1600&7\\
NGC7469&Sy1&0.0163&23:03:15.6&+08:52:26&900&450&450&8\\
VIIZW118&Sy1& 0.0803 &07:07:13.1&+64:35:59&4000&2000&2000&9\\
\hline
\end{tabular}
\caption{AGN observed using ISIS on the WHT.  Target co-ordinates taken from \textit{NASA/IPAC Extragalactic Database}. Spectral classifications taken from: 1. \protect \cite{Rodriguez-Ardila02}, 2. \protect \cite{Boroson92}, 3.\protect \cite{Constantin03}, 4. \protect \cite{Rudy00}, 5. \textit{NASA/IPAC Extragalactic Database}, 6. \protect \cite{Ryan07}, 7. \protect \cite{DeVaucouleurs75} 8. \protect \cite{Riffel06} 9. \protect \cite{Kunth79}.  Redshifts measured from the [SII]$\lambda\lambda 6717/32$ doublet (assumed to represent the systemic velocity of the galaxy), with heliocentric correction.  A typical error on these redshifts is $\sim$45km/s}
\label{ObjTable}
\end{table*}

\subsection{CCD Image Reduction}
\label{ImRed}

The raw CCD images first had the bias level subtracted, then they were  flat fielded and trimmed using the {\it IRAF} set of image reduction software.  A two dimensional dispersion correction and wavelength calibration function was derived from the calibration arc frames. This was applied to each science frame, again using standard {\it IRAF} routines.  The quality of the wavelength calibration was checked and corrected where necessary using the sky emission lines identified by \cite{Osterbrock96}. To extract the 1D spectrum from the 2D image, an extraction routine was written using the IDL programming language. We initially collapsed the 2D image along the direction of dispersion in order to obtain an indication of the spatial profile of the object within the slit.  This `average' profile was then fitted with a single Gaussian. We then returned to the original 2D image and extracted the 1D spectrum by summing the contribution of each pixel within $2\sigma$ limits of the average Gaussian profile, thus ensuring the same extraction window was used along the spectrum.  The underlying sky background was determined by extrapolation of the region between $3\sigma$ and $20\sigma$ either side of the object spectra.  These spectra were then flux calibrated using the standard star observations and the \textit{IRAF} routines \textit{noao.onedspec.standard, noao.onedspec.sensfunc} and \textit{noao.onedspec.calibrate}.  An attempt was then made to correct for telluric absorption by the division of a normalised, continuum removed (i.e. flat) spectrum of a standard star. However, since this introduces large uncertainties in the spectra between the wavelengths affected by telluric absorption features, we have not attempted to obtain strengths for coronal lines within these regions.

\subsection{General Corrections}

Due to the procedure used to obtain the required wavelength coverage, there are three associated spectra for each galaxy; blue (4400-5600\AA), red/short (5700-7000\AA) and red/long (7300-8600\AA), which we have joined together to form a single, almost continuous spectrum.  During the spectral joining process, it was noted that no significant corrections were needed to force the spectra covering the three wavelength ranges to align, thus giving us confidence in the flux calibration procedure and the photometric accuracy. All our spectra have been corrected for redshift, which was calculated from the average observed wavelength of the [S II]$\lambda$6717, 6732 doublet,  and geocentric velocities have been taken into account.  All velocities published here are  therefore relative to the [S II]$\lambda$6717, 6732 doublet, which we assume is representative of the host galaxy.  This is based on the fact that it has a low excitation potential and low critical density, meaning that it is likely to be formed away from the AGN. 

The spectra have been corrected for Galactic extinction using the wavelength dependent absorption function of \cite{Cardelli89} and the \cite{Schlegel98} $E(B-V)$ values obtained from NASA/IPAC Extragalactic Database. Due to the narrow slit widths used, we expect that the spectra suffer only minor  contamination from the host galaxy.  To test this assumption, we searched for the stellar absorption feature Mg b at $\lambda$5170\AA, but only found evidence of it in Mrk573 (the only Seyfert 2 in our sample), in which it has an equivalent width of $\sim$5\AA \ . Because we concentrate on the profiles and relative strengths of emission lines at wavelengths which are largely unaffected by the stellar features, we have not corrected for this contamination. 

\begin{figure*}
\begin{center}
	\includegraphics[width=18cm, height =
	12cm]{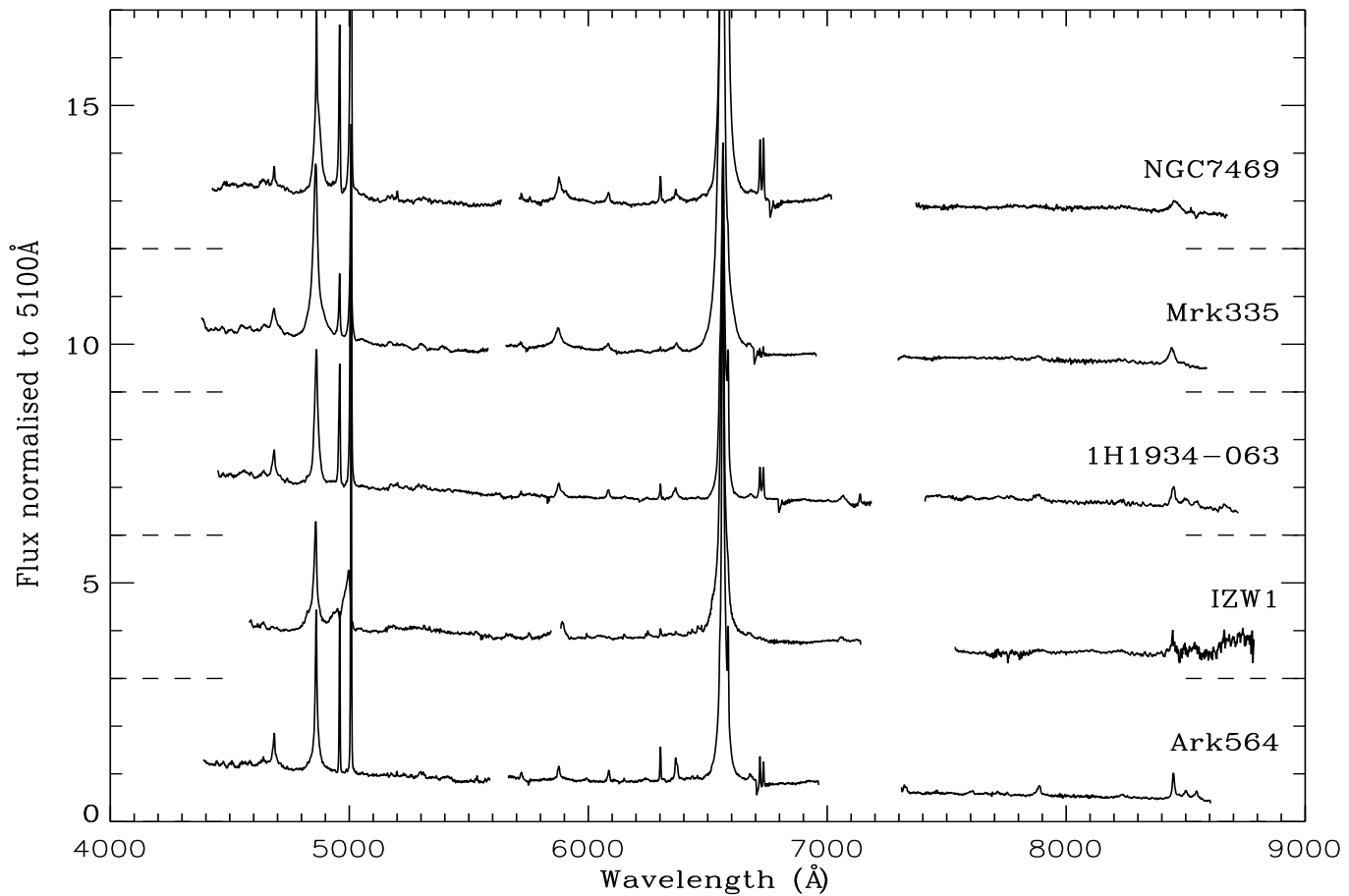}
\end{center}
\caption{Flux calibrated, redshift corrected, FeII subtracted spectra.  Flux has been normalised to the level at 5100\AA and offset for clarity.  The dotted lines either side of the plot show the level of the offset.  For a clearer depiction of the lines discussed here, we refer the reader to Figs. \ref{H_a} - \ref{FeXIV}}
\label{FullSpec1}
\end{figure*}

\begin{figure*}
\begin{center}
	\includegraphics[width=18cm, height =
	12cm]{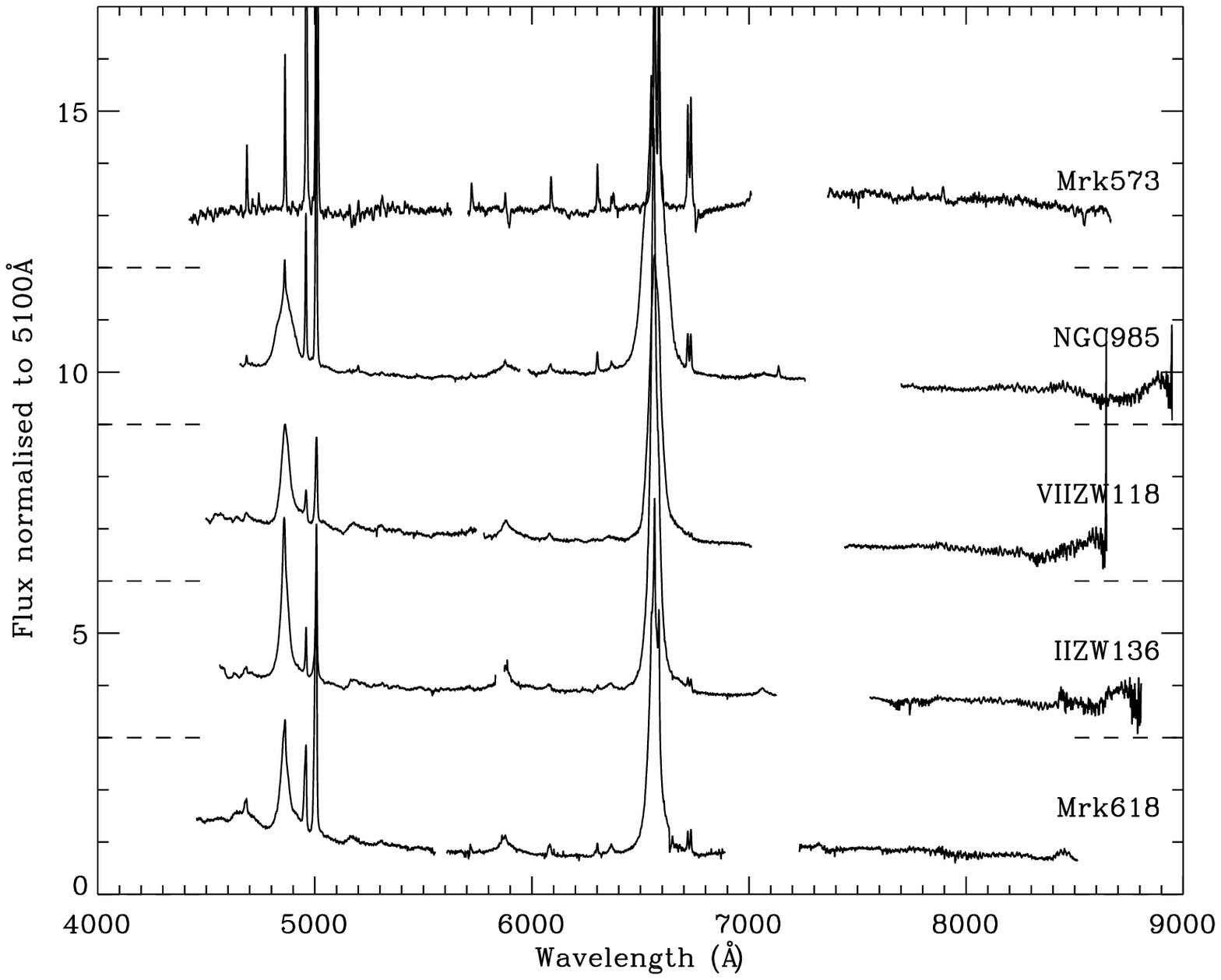}
\end{center}
\contcaption{}
\label{FullSpec2}
\end{figure*}

\subsection{Continuum Fitting}

Since we wish to measure the total flux and profile of the emission lines, it was necessary to accurately fit and subtract the underlying continuum. For blends of weak lines, this can be problematic as it is  difficult to distinguish between the continuum and  extended line wings, thereby making the continuum fit somewhat subjective. In an attempt to more objectively determine the continuum levels, we modelled the whole spectrum with a gently inflecting fourth order polynomial fitted to regions of the spectrum that were perceived by eye to be free of emission features.  Each fit was then inspected at the wavelength of the lines of interest to check that it was  a sensible fit to the local underlying continuum.  As some of the weaker emission lines lie on the broad wings of the H$\alpha$ profile, we first fit the H$\alpha$ line and remove the contribution of the broad wings from the weak lines. 

\subsection{Fe II Lines}
Permitted emission from Fe II is often detected in the optical spectra of AGN, and is particularly strong in many NLS1s, in which multiplets of  broad lines blend together to form a pseudo-continuum.  As some of our lines of interest lie close to strong Fe II features, we decided to remove them by using a method closely based on that outlined in \cite{Boroson92}.  In their study of the strong Fe II lines in the spectrum of IZW1, \cite{Veron-Cetty04} show that this emission contains two velocity components.  To account for this we used their published line lists as a template, and broadened the two sets of lines by convolving this template with two Gaussians with independent variables of velocity width and intensity.  In their method, \cite{Boroson92} produced a matrix of Fe II spectra, from which they selected the one that most closely matched the Fe II emission in a particular object's spectrum.  Increasing computing processing speeds since their original  work fortunately means that we can now produce the broadened Fe II spectrum `on the fly' using an IDL based GUI \textit{REMOVEFEII} \footnote[1]{Available from http://www.dur.ac.uk/j.r.mullaney}. This enables much more control over the range of input variables, making the Fe II feature removal a more dynamic and controlled process.  As was noted by \cite{Landt07}, it is difficult to determine the true widths of the Fe II lines as above a certain velocity width the lines blend together into a single broad feature.  To counter this it was noted that by fitting a single line that was unique to either the narrow or broad component templates and that was not part of an Fe II blend, a reasonably independent estimate of the strength and width of the two templates could be determined.  A limitation to this method is that it is dependent on the assumption that all our AGN display the same Fe II emission lines, with the same relative strengths, as found in IZW1.

In practice it was found that, in general, this procedure was able to remove a significant proportion of the narrow Fe II emission component, but it was difficult to fully account for the broad component of the emission lines.  A qualitative assessment of the line profiles affected by Fe II emission is therefore possible, but the level of uncertainty does not allow reliable measurements in cases where the FeII blends are strong -between $\lambda\lambda$4100-4600 and $\lambda\lambda$5150 -5400 (in particular the region of [Fe XIV]$\lambda$5303).

\subsection{Line fitting}
All the emission lines discussed in detail in the subsequent sections were modelled by fitting multiple Gaussian profiles, which we argue is more statistically robust than the Lorentzian component fitting method used in some previous NLS1 studies (see our appendix; \citealt{Goncalves99}, \citealt{Veron01}).  Blends (excluding the Fe II lines discussed above) were dealt with by simultaneously fitting the blended lines.  This method worked well for the majority of blended lines.  In regions of the spectra where broad lines are blended together with narrow lines, (in particular the H$\alpha$ and [NII]$\lambda\lambda$6548,6584 lines), once we had fitted these lines we compared the widths of the blended narrow lines with those of the unblended narrow [OIII]$\lambda\lambda$4959,5007 lines as a check that the fit that been performed correctly.

The 90\% confidence errors associated with the parameters derived for each fitted Gaussian (i.e. intensity, width and central wavelength) were determined by deviating each of the parameters from their optimum values until a change in $\chi^2$ equal to the published 90\% confidence level for the appropriate number of free fitting parameters was obtained. In addition to these random errors, we found that uncertainties in the continuum model leads to significant systematic errors for the line parameters (especially the broadest components).  To estimate this, we rescaled the continuum by $\pm$1$\sigma$ (adopting $RMS$ as 1$\sigma$) and refit the line models, thereby quantifying the impact of continuum levels that are set too high or too low.

The intensities, widths and shifts of the multiple Gaussian fits for the lines discussed in this paper, along with their associated random and systematic errors, are listed in Tables \ref{inten_sep_table} - \ref{pos_sep_table}, which we discuss in turn below.  We note that, in Table \ref{width_sep_table} we have not corrected for instrumental broadening as none of the components of the lines of interest had widths close to the instrumental broadening determined from the sky lines ($\sim$2\AA).  Therefore, when considered in quadrature, instrumental broadening would have an insignificant effect on the measured line widths. 

\section{Emission Line Results}

\subsection{The Broad Emission Lines}
Due to the clear distinction between the spectrum of the only Seyfert 2 in our sample, Mrk573, and all the other galaxies we observed (Seyfert 1s, with a range of broad emission line widths), we will discuss this galaxy in a separate section.

\subsubsection{Balmer Lines}

\begin{table*}
\centering
\caption{Modelled line flux, as a percentage of the H$\beta$ line.  Where no value is listed, there was no significant indication of the presence of the line or component.  The $\pm$ errors refer to the 90\% confidence interval of the measurement, the (\%) errors refer to the systematic errors associated with the estimation of the continuum flux levels.  The H$\beta$ line flux has units of 10$^{-13}$ ergs/s/cm$^{2}$}
\label{inten_sep_table}
\begin{tabular}{@{}lcccccccccc@{}}
\hline
Line&Ark564&IZW1&1H1934-063&Mrk335&NGC7469\\
\hline
H$\beta$ Flux&2.76&2.62&6.11&6.70&5.43\\
\hline
H$_{\alpha}$$\lambda$6563&454.2 $\pm$ 5.0 (5\%)&453.8 $\pm$ 3.3 (15\%)&320.4 $\pm$ 1.5 (5\%)&350.1 $\pm$ 2.4 ($<$2.5\%)&487.5 $\pm$ 3.2 (5\%)\\
\hspace{0.2cm}Broad&184.9 $\pm$ 2.5 ($<$2.5\%)&228.0 $\pm$ 1.9 (10\%)&76.06 $\pm$ 0.78 ($<$2.5\%)&155.0 $\pm$ 1.4 ($<$2.5\%)&424.1 $\pm$ 2.8 (5\%)\\
\hspace{0.2cm}Intermediate&169.1 $\pm$ 2.1 (10\%)&175.5 $\pm$ 1.4 (5\%)&198.03 $\pm$ 0.86 (5\%)&139.4 $\pm$ 1.1 ($<$2.5\%)&-\\
\hspace{0.2cm}Narrow&100.2 $\pm$ 1.3 (5\%)&50.33 $\pm$ 0.59 (50\%)&46.27 $\pm$ 0.38 ($<$2.5\%)&55.77 $\pm$ 0.66 (5\%)&63.41 $\pm$ 0.64 ($<$2.5\%)\\
H$_{\beta}$$\lambda$4861&100.0 $\pm$ 1.4 (10\%)&100.00 $\pm$ 0.92 (10\%)&100.00 $\pm$ 0.48 (15\%)&100.00 $\pm$ 0.74 (5\%)&100.00 $\pm$ 0.83 (5\%)\\
\hspace{0.2cm}Broad&46.80 $\pm$ 0.92 (5\%)&61.66 $\pm$ 0.69 (15\%)&43.10 $\pm$ 0.30 (10\%)&60.16 $\pm$ 0.52 (5\%)&85.76 $\pm$ 0.74 (5\%)\\
\hspace{0.2cm}Intermediate&36.36 $\pm$ 0.58 (10\%)&28.18 $\pm$ 0.33 (5\%)&50.75 $\pm$ 0.25 (20\%)&28.92 $\pm$ 0.30 (5\%)&-\\
\hspace{0.2cm}Narrow&16.84 $\pm$ 0.33 (20\%)&10.16 $\pm$ 0.19 (10\%)&6.15 $\pm$ 0.10 (15\%)&10.92 $\pm$ 0.20 (5\%)&14.24 $\pm$ 0.24 ($<$2.5\%)\\
$[$O III$]$$\lambda$5007&88.50 $\pm$ 0.96 (5\%)&65.10 $\pm$ 0.61 (15\%)&38.87 $\pm$ 0.21 ($<$2.5\%)&29.32 $\pm$ 0.25 (5\%)&111.13 $\pm$ 0.77 ($<$2.5\%)\\
\hspace{0.2cm}Broad&44.02 $\pm$ 0.54 ($<$2.5\%)&48.66 $\pm$ 0.49 (15\%)&28.19 $\pm$ 0.17 ($<$2.5\%)&18.19 $\pm$ 0.20 (5\%)&46.32 $\pm$ 0.44 (5\%)\\
\hspace{0.2cm}Narrow&44.48 $\pm$ 0.51 (5\%)&16.45 $\pm$ 0.26 (10\%)&10.679 $\pm$ 0.098 ($<$2.5\%)&11.13 $\pm$ 0.12 (5\%)&64.80 $\pm$ 0.45 ($<$2.5\%)\\
He I$\lambda$5876&17.0 $\pm$ 1.1 (45\%)&-&12.22 $\pm$ 0.34 (35\%)&23.27 $\pm$ 0.81 (30\%)&23.33 $\pm$ 0.89 (35\%)\\
\hspace{0.2cm}Broad&12.81 $\pm$ 0.99 (60\%)&-&8.59 $\pm$ 0.30 (40\%)&16.45 $\pm$ 0.74 (35\%)&18.05 $\pm$ 0.80 (40\%)\\
\hspace{0.2cm}Narrow&4.15 $\pm$ 0.37 ($<$2.5\%)&-&3.63 $\pm$ 0.17 (25\%)&6.82 $\pm$ 0.32 (15\%)&5.27 $\pm$ 0.38 (10\%)\\
He II$\lambda$4686&25.54 $\pm$ 0.64 (15\%)&-&22.52 $\pm$ 0.24 (15\%)&12.18 $\pm$ 0.29 (10\%)&10.71 $\pm$ 0.39 (40\%)\\
\hspace{0.2cm}Broad&18.97 $\pm$ 0.57 (10\%)&-&15.26 $\pm$ 0.21 (5\%)&8.07 $\pm$ 0.25 (5\%)&8.83 $\pm$ 0.36 (45\%)\\
\hspace{0.2cm}Narrow&6.57 $\pm$ 0.25 (25\%)&-&7.25 $\pm$ 0.11 (35\%)&4.11 $\pm$ 0.15 (20\%)&1.88 $\pm$ 0.15 (5\%)\\
$[$Fe VII$]$$\lambda$6087&3.95 $\pm$ 0.29 (25\%)&-&3.44 $\pm$ 0.14 (30\%)&3.12 $\pm$ 0.26 (65\%)&5.35 $\pm$ 0.66 (40\%)\\
\hspace{0.2cm}Broad&3.95 $\pm$ 0.29 (25\%)&-&3.44 $\pm$ 0.14 (30\%)&3.12 $\pm$ 0.26 (65\%)&2.79 $\pm$ 0.55 (50\%)\\
\hspace{0.2cm}Narrow&-&-&-&-&2.56 $\pm$ 0.35 (30\%)\\
$[$Fe X$]$$\lambda$6374&9.36 $\pm$ 0.66 (25\%)&-&6.26 $\pm$ 0.21 (25\%)&5.68 $\pm$ 0.59 (55\%)&7.29 $\pm$ 0.68 (70\%)\\
\hspace{0.2cm}Broad&8.03 $\pm$ 0.56 (25\%)&-&6.26 $\pm$ 0.21 (25\%)&4.80 $\pm$ 0.54 (60\%)&7.29 $\pm$ 0.68 (70\%)\\
\hspace{0.2cm}Narrow&1.32 $\pm$ 0.33 (25\%)&-&-&0.88 $\pm$ 0.24 (15\%)&-\\
$[$Fe XI$]$$\lambda$7892&8.83 $\pm$ 0.57 (30\%)&-&7.80 $\pm$ 0.23 (35\%)&2.24 $\pm$ 0.17 (60\%)&-\\
\hspace{0.2cm}Broad&5.96 $\pm$ 0.49 (30\%)&-&7.80 $\pm$ 0.23 (35\%)&2.24 $\pm$ 0.17 (60\%)&-\\
\hspace{0.2cm}Narrow&2.88 $\pm$ 0.30 (35\%)&-&-&-&-\\
$[$Fe XIV$]$$\lambda$5303&1.81 $\pm$ 0.33 (35\%)&-&1.81 $\pm$ 0.11 (60\%)&-&-\\
\hspace{0.2cm}Broad&1.81 $\pm$ 0.33 (35\%)&-&1.81 $\pm$ 0.11 (60\%)&-&-\\
\hline
\end{tabular}

\begin{tabular}{@{}lcccccccccc@{}}
\hline
Line&Mrk618&IIZW136&VIIZW118&NGC985&Mrk573\\
\hline
H$\beta$ Flux&1.84&4.64&1.47&3.63&0.400\\
\hline
H$_{\alpha}$$\lambda$6563&331.1 $\pm$ 3.4 (5\%)&383.0 $\pm$ 2.3 (5\%)&412.8 $\pm$ 3.4 (10\%)&527.5 $\pm$ 1.9 ($<$2.5\%)&382. $\pm$ 10.0 ($<$2.5\%)\\
\hspace{0.2cm}Broad&184.1 $\pm$ 2.2 (5\%)&116.2 $\pm$ 1.00 ($<$2.5\%)&122.0 $\pm$ 1.6 (10\%)&501.4 $\pm$ 1.8 ($<$2.5\%)&-\\
\hspace{0.2cm}Intermediate&136.5 $\pm$ 1.6 (10\%)&209.7 $\pm$ 1.3 (5\%)&-&-&242.8 $\pm$ 6.8 ($<$2.5\%)\\
\hspace{0.2cm}Narrow&10.54 $\pm$ 0.42 ($<$2.5\%)&57.15 $\pm$ 0.47 ($<$2.5\%)&290.8 $\pm$ 2.3 (10\%)&26.09 $\pm$ 0.32 (5\%)&138.9 $\pm$ 4.0 ($<$2.5\%)\\
H$_{\beta}$$\lambda$4861&100.0 $\pm$ 1.2 (20\%)&100.00 $\pm$ 0.76 (5\%)&100.0 $\pm$ 1.0 (5\%)&100.00 $\pm$ 0.46 (10\%)&100.0 $\pm$ 3.3 (10\%)\\
\hspace{0.2cm}Broad&76.37 $\pm$ 0.95 (10\%)&34.50 $\pm$ 0.46 (5\%)&41.11 $\pm$ 0.66 (10\%)&95.07 $\pm$ 0.43 (10\%)&-\\
\hspace{0.2cm}Intermediate&22.03 $\pm$ 0.48 (45\%)&55.84 $\pm$ 0.42 (10\%)&-&-&66.0 $\pm$ 2.5 (5\%)\\
\hspace{0.2cm}Narrow&1.61 $\pm$ 0.24 (40\%)&9.66 $\pm$ 0.17 ($<$2.5\%)&58.89 $\pm$ 0.60 (5\%)&4.926 $\pm$ 0.095 (10\%)&34.0 $\pm$ 1.4 (15\%)\\
$[$O III$]$$\lambda$5007&80.54 $\pm$ 0.78 (5\%)&23.94 $\pm$ 0.26 (5\%)&20.54 $\pm$ 0.37 (10\%)&60.36 $\pm$ 0.27 ($<$2.5\%)&1216. $\pm$ 28. ($<$2.5\%)\\
\hspace{0.2cm}Broad&65.06 $\pm$ 0.64 (5\%)&12.97 $\pm$ 0.21 (5\%)&11.08 $\pm$ 0.28 (10\%)&25.85 $\pm$ 0.18 ($<$2.5\%)&785. $\pm$ 18. ($<$2.5\%)\\
\hspace{0.2cm}Narrow&15.48 $\pm$ 0.23 ($<$2.5\%)&10.97 $\pm$ 0.12 (5\%)&9.46 $\pm$ 0.22 (5\%)&34.51 $\pm$ 0.15 ($<$2.5\%)&431. $\pm$ 10. ($<$2.5\%)\\
He I$\lambda$5876&26.0 $\pm$ 1.1 (35\%)&-&25.4 $\pm$ 1.1 (20\%)&15.68 $\pm$ 0.62 (40\%)&9.1 $\pm$ 1.4 (45\%)\\
\hspace{0.2cm}Broad&18.8 $\pm$ 1.0 (40\%)&-&17.48 $\pm$ 0.94 (25\%)&14.77 $\pm$ 0.59 (40\%)&9.1 $\pm$ 1.4 (45\%)\\
\hspace{0.2cm}Narrow&7.22 $\pm$ 0.53 (25\%)&-&7.94 $\pm$ 0.51 (5\%)&0.91 $\pm$ 0.18 (15\%)&-\\
He II$\lambda$4686&18.83 $\pm$ 0.63 (20\%)&5.00 $\pm$ 0.24 (45\%)&5.28 $\pm$ 0.39 (35\%)&1.145 $\pm$ 0.050 (25\%)&35.6 $\pm$ 1.9 (20\%)\\
\hspace{0.2cm}Broad&13.51 $\pm$ 0.57 (25\%)&3.00 $\pm$ 0.21 (65\%)&3.16 $\pm$ 0.33 (55\%)&1.145 $\pm$ 0.050 (25\%)&22.0 $\pm$ 1.5 (25\%)\\
\hspace{0.2cm}Narrow&5.32 $\pm$ 0.26 (5\%)&2.00 $\pm$ 0.11 (20\%)&2.12 $\pm$ 0.21 (10\%)&-&13.7 $\pm$ 1.0 (10\%)\\
$[$Fe VII$]$$\lambda$6087&4.29 $\pm$ 0.30 (35\%)&4.27 $\pm$ 0.26 (40\%)&2.81 $\pm$ 0.31 (40\%)&2.80 $\pm$ 0.15 (50\%)&28.0 $\pm$ 1.9 (20\%)\\
\hspace{0.2cm}Broad&4.29 $\pm$ 0.30 (35\%)&4.27 $\pm$ 0.26 (40\%)&2.81 $\pm$ 0.31 (40\%)&2.80 $\pm$ 0.15 (50\%)&28.0 $\pm$ 1.9 (20\%)\\
\hspace{0.2cm}Narrow&-&-&-&-&-\\
$[$Fe X$]$$\lambda$6374&6.60 $\pm$ 0.45 (30\%)&6.47 $\pm$ 0.32 (45\%)&5.57 $\pm$ 0.59 (80\%)&-&15.4 $\pm$ 2.1 (30\%)\\
\hspace{0.2cm}Broad&6.60 $\pm$ 0.45 (30\%)&6.47 $\pm$ 0.32 (45\%)&5.57 $\pm$ 0.59 (80\%)&-&15.4 $\pm$ 2.1 (30\%)\\
\hspace{0.2cm}Narrow&-&-&-&-&-\\
$[$Fe XI$]$$\lambda$7892&-&-&-&-&11.1 $\pm$ 1.4 (80\%)\\
\hspace{0.2cm}Broad&-&-&-&-&11.1 $\pm$ 1.4 (80\%)\\
\hspace{0.2cm}Narrow&-&-&-&-&-\\
$[$Fe XIV$]$$\lambda$5303&-&-&-&-&-\\
\hspace{0.2cm}Broad&-&-&-&-&-\\
\hline
\end{tabular}

\end{table*}

\begin{table*}
\centering
\caption{Line widths in km/s.  As above, the random errors are the 90\% confidence interval and the systematic uncertainties arising from the continuum model are given in parenthesis.
measurement.}
\label{width_sep_table}
\begin{tabular}{@{}lcccccccccc@{}}
\hline
Line&Ark564&IZW1&1H1934-063&Mrk335&NGC7469\\
\hline
H$_{\alpha}$$\lambda$6563& & & & & \\
\hspace{0.2cm}Broad&2578. $\pm$ 24. (10\%)&3247. $\pm$ 16. (20\%)&2837. $\pm$ 26. (10\%)&3920. $\pm$ 27. (5\%)&2066.1 $\pm$ 6.4 ($<$2.5\%)\\
\hspace{0.2cm}Intermediate&963.1 $\pm$ 7.3 (5\%)&918.3 $\pm$ 4.1 (25\%)&1271.7 $\pm$ 3.7 ($<$2.5\%)&1333.4 $\pm$ 7.3 ($<$2.5\%)&-\\
\hspace{0.2cm}Narrow&320.9 $\pm$ 3.2 ($<$2.5\%)&328.2 $\pm$ 3.2 (20\%)&411.4 $\pm$ 3.2 ($<$2.5\%)&707.1 $\pm$ 8.2 ($<$2.5\%)&230.4 $\pm$ 2.3 ($<$2.5\%)\\
H$_{\beta}$$\lambda$4861& & & & & \\
\hspace{0.2cm}Broad&3175. $\pm$ 60. (25\%)&4049. $\pm$ 43. (15\%)&2624. $\pm$ 18. (25\%)&3838. $\pm$ 29. ($<$2.5\%)&2046. $\pm$ 15. (5\%)\\
\hspace{0.2cm}Intermediate&897. $\pm$ 12. (15\%)&967. $\pm$ 11. (5\%)&1059.7 $\pm$ 4.3 (10\%)&1292. $\pm$ 12. ($<$2.5\%)&-\\
\hspace{0.2cm}Narrow&290.1 $\pm$ 5.6 (10\%)&388.8 $\pm$ 7.4 (5\%)&317.8 $\pm$ 6.2 (10\%)&685. $\pm$ 14. ($<$2.5\%)&329.6 $\pm$ 6.2 ($<$2.5\%)\\
$[$O III$]$$\lambda$5007& & & & & \\
\hspace{0.2cm}Broad&336.1 $\pm$ 2.4 (5\%)&2044. $\pm$ 20. (5\%)&538.0 $\pm$ 3.0 ($<$2.5\%)&893. $\pm$ 11. (5\%)&744.2 $\pm$ 6.6 (10\%)\\
\hspace{0.2cm}Narrow&161.2 $\pm$ 1.2 ($<$2.5\%)&759. $\pm$ 13. (5\%)&191.1 $\pm$ 1.8 ($<$2.5\%)&246.3 $\pm$ 2.4 (5\%)&284.6 $\pm$ 1.2 ($<$2.5\%)\\
He I$\lambda$5876& & & & & \\
\hspace{0.2cm}Broad&3250 $\pm$ 510 (25\%)&-&1852. $\pm$ 72. (45\%)&6970 $\pm$ 380 (30\%)&2910 $\pm$ 160 (25\%)\\
\hspace{0.2cm}Narrow&475. $\pm$ 46. ($<$2.5\%)&-&576. $\pm$ 31. (15\%)&1336. $\pm$ 76. (10\%)&665. $\pm$ 57. (5\%)\\
He II$\lambda$4686& & & & & \\
\hspace{0.2cm}Broad&2305. $\pm$ 76. (20\%)&-&2535. $\pm$ 37. (20\%)&2666. $\pm$ 93. (5\%)&1700 $\pm$ 110 (25\%)\\
\hspace{0.2cm}Narrow&474. $\pm$ 22. (20\%)&-&686. $\pm$ 12. (20\%)&957. $\pm$ 40. (10\%)&290. $\pm$ 25. (5\%)\\
$[$Fe VII$]$$\lambda$6087& & & & & \\
\hspace{0.2cm}Broad&446. $\pm$ 37. (15\%)&-&572. $\pm$ 27. (20\%)&1380 $\pm$ 140 (45\%)&1270 $\pm$ 360 (45\%)\\
\hspace{0.2cm}Narrow&-&-&-&-&518. $\pm$ 85. (20\%)\\
$[$Fe X$]$$\lambda$6374& & & & & \\
\hspace{0.2cm}Broad&749. $\pm$ 62. (25\%)&-&1127. $\pm$ 42. (15\%)&2680 $\pm$ 420 (35\%)&2560 $\pm$ 300 (35\%)\\
\hspace{0.2cm}Narrow&268. $\pm$ 73. (10\%)&-&-&510 $\pm$ 170 (10\%)&-\\
$[$Fe XI$]$$\lambda$7892& & & & & \\
\hspace{0.2cm}Broad&1140 $\pm$ 110 (40\%)&-&1457. $\pm$ 48. (20\%)&1330 $\pm$ 120 (35\%)&-\\
\hspace{0.2cm}Narrow&428. $\pm$ 51. (20\%)&-&-&-&-\\
$[$Fe XIV$]$$\lambda$5303& & & & & \\
\hspace{0.2cm}Broad&540 $\pm$ 110 (30\%)&-&623. $\pm$ 55. (35\%)&-&-\\
\hline
\end{tabular}

\begin{tabular}{@{}lcccccccccc@{}}
\hline
Line&Mrk618&IIZW136&VIIZW118&NGC985&Mrk573\\
\hline
H$_{\alpha}$$\lambda$6563& & & & & \\
\hspace{0.2cm}Broad&3221. $\pm$ 26. (5\%)&4474. $\pm$ 30. (5\%)&5242. $\pm$ 59. (20\%)&4987.5 $\pm$ 7.8 ($<$2.5\%)&-\\
\hspace{0.2cm}Intermediate&1681. $\pm$ 15. ($<$2.5\%)&2093.1 $\pm$ 5.9 ($<$2.5\%)&-&-&533.9 $\pm$ 9.1 ($<$2.5\%)\\
\hspace{0.2cm}Narrow&191.1 $\pm$ 8.2 ($<$2.5\%)&675.1 $\pm$ 4.6 ($<$2.5\%)&2503.1 $\pm$ 8.7 ($<$2.5\%)&519.3 $\pm$ 8.7 (5\%)&180.1 $\pm$ 3.7 ($<$2.5\%)\\
H$_{\beta}$$\lambda$4861& & & & & \\
\hspace{0.2cm}Broad&3343. $\pm$ 35. (10\%)&5130. $\pm$ 68. (20\%)&4543. $\pm$ 70. (20\%)&5003. $\pm$ 18. (5\%)&-\\
\hspace{0.2cm}Intermediate&1360. $\pm$ 33. (30\%)&2353. $\pm$ 14. (5\%)&-&-&546. $\pm$ 19. (20\%)\\
\hspace{0.2cm}Narrow&240. $\pm$ 64. (5\%)&649. $\pm$ 12. ($<$2.5\%)&2293. $\pm$ 18. ($<$2.5\%)&467. $\pm$ 12. (10\%)&185.8 $\pm$ 7.4 (5\%)\\
$[$O III$]$$\lambda$5007& & & & & \\
\hspace{0.2cm}Broad&810.1 $\pm$ 4.8 (5\%)&1379. $\pm$ 26. (15\%)&772. $\pm$ 22. (10\%)&803.5 $\pm$ 5.4 (5\%)&523.1 $\pm$ 2.4 ($<$2.5\%)\\
\hspace{0.2cm}Narrow&267.2 $\pm$ 3.6 ($<$2.5\%)&339.1 $\pm$ 3.6 (5\%)&442. $\pm$ 11. ($<$2.5\%)&309.8 $\pm$ 1.2 ($<$2.5\%)&169.6 $\pm$ 1.2 ($<$2.5\%)\\
He I$\lambda$5876& & & & & \\
\hspace{0.2cm}Broad&5500 $\pm$ 340 (35\%)&-&5470 $\pm$ 360 (15\%)&5150 $\pm$ 280 (25\%)&254. $\pm$ 45. (30\%)\\
\hspace{0.2cm}Narrow&1600 $\pm$ 130 (10\%)&-&1810 $\pm$ 140 ($<$2.5\%)&520 $\pm$ 130 (10\%)&-\\
He II$\lambda$4686& & & & & \\
\hspace{0.2cm}Broad&4110 $\pm$ 190 (10\%)&2990 $\pm$ 270 (45\%)&2070 $\pm$ 260 (25\%)&423. $\pm$ 22. (15\%)&430. $\pm$ 30. (20\%)\\
\hspace{0.2cm}Narrow&920. $\pm$ 50. (5\%)&1061. $\pm$ 68. (5\%)&940 $\pm$ 100 (10\%)&-&198. $\pm$ 17. (5\%)\\
$[$Fe VII$]$$\lambda$6087& & & & & \\
\hspace{0.2cm}Broad&775. $\pm$ 63. (15\%)&2020 $\pm$ 190 (20\%)&1060 $\pm$ 140 (25\%)&980. $\pm$ 65. (35\%)&378. $\pm$ 30. (15\%)\\
\hspace{0.2cm}Narrow&-&-&-&-&-\\
$[$Fe X$]$$\lambda$6374& & & & & \\
\hspace{0.2cm}Broad&1290 $\pm$ 130 (15\%)&2740 $\pm$ 180 (25\%)&2980 $\pm$ 400 (45\%)&-&387. $\pm$ 58. (20\%)\\
\hspace{0.2cm}Narrow&-&-&-&-&-\\
$[$Fe XI$]$$\lambda$7892& & & & & \\
\hspace{0.2cm}Broad&-&-&-&-&330. $\pm$ 43. (45\%)\\
\hspace{0.2cm}Narrow&-&-&-&-&-\\
$[$Fe XIV$]$$\lambda$5303& & & & & \\
\hspace{0.2cm}Broad&-&-&-&-&-\\
\hline
\end{tabular}

\end{table*}

\begin{table*}
\centering
\caption{Line shift from systemic velocity shift (in km/s), as determined from the [S II]$\lambda$6718 doublet.  We use the convention that negative numbers indicate a net redshift.}
\label{pos_sep_table}
\begin{tabular}{@{}lcccccccccc@{}}
\hline
Line&Ark564&IZW1&1H1934-063&Mrk335&NGC7469\\
\hline
H$_{\alpha}$$\lambda$6563& & & & & \\
\hspace{0.2cm}Broad&-71. $\pm$ 13.&234.4 $\pm$ 8.2&-108. $\pm$ 14.&118. $\pm$ 14.&-249.2 $\pm$ 0.00\\
\hspace{0.2cm}Intermediate&-32.4 $\pm$ 4.1&99.6 $\pm$ 2.3&10.5 $\pm$ 1.8&298.3 $\pm$ 0.00&-\\
\hspace{0.2cm}Narrow&-7.3 $\pm$ 1.8&-19.2 $\pm$ 1.8&-0.5 $\pm$ 1.8&-8.4 $\pm$ 4.6&-81.5 $\pm$ 0.00\\
H$_{\beta}$$\lambda$4861& & & & & \\
\hspace{0.2cm}Broad&-35. $\pm$ 31.&461. $\pm$ 21.&-76.3 $\pm$ 9.9&90. $\pm$ 12.&-312.7 $\pm$ 6.2\\
\hspace{0.2cm}Intermediate&-7.3 $\pm$ 6.8&252.1 $\pm$ 5.6&0.8 $\pm$ 2.5&257.5 $\pm$ 6.2&-\\
\hspace{0.2cm}Narrow&-10.4 $\pm$ 3.1&72.0 $\pm$ 3.7&-63.9 $\pm$ 3.1&37.9 $\pm$ 6.2&-91.9 $\pm$ 0.00\\
$[$O III$]$$\lambda$5007& & & & & \\
\hspace{0.2cm}Broad&13.6 $\pm$ 1.8&1375.8 $\pm$ 9.6&105.9 $\pm$ 1.8&149.0 $\pm$ 5.4&215.5 $\pm$ 3.6\\
\hspace{0.2cm}Narrow&-42.12 $\pm$ 0.60&537.2 $\pm$ 6.6&-37.9 $\pm$ 1.2&-15.1 $\pm$ 1.2&-20.44 $\pm$ 0.60\\
He I$\lambda$5876& & & & & \\
\hspace{0.2cm}Broad&-220 $\pm$ 180&-&-369. $\pm$ 38.&-230 $\pm$ 190&-754. $\pm$ 79.\\
\hspace{0.2cm}Narrow&-8. $\pm$ 25.&-&52. $\pm$ 16.&141. $\pm$ 37.&-81. $\pm$ 29.\\
He II$\lambda$4686& & & & & \\
\hspace{0.2cm}Broad&-23. $\pm$ 38.&-&30. $\pm$ 19.&-428. $\pm$ 45.&59. $\pm$ 45.\\
\hspace{0.2cm}Narrow&55. $\pm$ 11.&-&133.5 $\pm$ 6.4&161. $\pm$ 19.&35. $\pm$ 13.\\
$[$Fe VII$]$$\lambda$6087& & & & & \\
\hspace{0.2cm}Broad&73. $\pm$ 19.&-&179. $\pm$ 14.&359. $\pm$ 69.&520 $\pm$ 150\\
\hspace{0.2cm}Narrow&-&-&-&-&99. $\pm$ 39.\\
$[$Fe X$]$$\lambda$6374& & & & & \\
\hspace{0.2cm}Broad&256. $\pm$ 31.&-&510. $\pm$ 19.&450 $\pm$ 190&440 $\pm$ 150\\
\hspace{0.2cm}Narrow&86. $\pm$ 41.&-&-&228. $\pm$ 83.&-\\
$[$Fe XI$]$$\lambda$7892& & & & & \\
\hspace{0.2cm}Broad&424. $\pm$ 55.&-&396. $\pm$ 23.&473. $\pm$ 60.&-\\
\hspace{0.2cm}Narrow&170. $\pm$ 26.&-&-&-&-\\
$[$Fe XIV$]$$\lambda$5303& & & & & \\
\hspace{0.2cm}Broad&425. $\pm$ 59.&-&720. $\pm$ 24.&-&-\\
\hline
\end{tabular}

\begin{tabular}{@{}lcccccccccc@{}}
\hline
Line&Mrk618&IIZW136&VIIZW118&NGC985&Mrk573\\
\hline
H$_{\alpha}$$\lambda$6563& & & & & \\
\hspace{0.2cm}Broad&20. $\pm$ 14.&32. $\pm$ 16.&-553. $\pm$ 32.&-88.9 $\pm$ 4.6&-\\
\hspace{0.2cm}Intermediate&-5.1 $\pm$ 8.2&-29.9 $\pm$ 3.2&-&-&-137.5 $\pm$ 4.6\\
\hspace{0.2cm}Narrow&-96.9 $\pm$ 4.6&177.1 $\pm$ 2.3&-186.6 $\pm$ 5.0&-20.4 $\pm$ 4.1&-57.6 $\pm$ 0.00\\
H$_{\beta}$$\lambda$4861& & & & & \\
\hspace{0.2cm}Broad&125. $\pm$ 22.&525. $\pm$ 35.&-704. $\pm$ 38.&-195.0 $\pm$ 9.3&-\\
\hspace{0.2cm}Intermediate&85. $\pm$ 17.&44.1 $\pm$ 7.4&-&-&-135.1 $\pm$ 6.2\\
\hspace{0.2cm}Narrow&-84. $\pm$ 54.&189.0 $\pm$ 6.8&-126.5 $\pm$ 9.9&-16.7 $\pm$ 5.6&-100.0 $\pm$ 0.00\\
$[$O III$]$$\lambda$5007& & & & & \\
\hspace{0.2cm}Broad&235.2 $\pm$ 2.4&222. $\pm$ 12.&106. $\pm$ 11.&104.1 $\pm$ 0.00&-131.2 $\pm$ 1.2\\
\hspace{0.2cm}Narrow&-49.3 $\pm$ 2.4&-0.3 $\pm$ 2.4&-89.3 $\pm$ 6.0&9.4 $\pm$ 0.00&-69.51 $\pm$ 0.60\\
He I$\lambda$5876& & & & & \\
\hspace{0.2cm}Broad&-410 $\pm$ 170&-&-1520 $\pm$ 170&-260 $\pm$ 130&-61. $\pm$ 23.\\
\hspace{0.2cm}Narrow&202. $\pm$ 68.&-&-146. $\pm$ 68.&-27. $\pm$ 62.&-\\
He II$\lambda$4686& & & & & \\
\hspace{0.2cm}Broad&-651. $\pm$ 97.&-870 $\pm$ 130&-920 $\pm$ 130&21.2 $\pm$ 6.4&-109. $\pm$ 17.\\
\hspace{0.2cm}Narrow&261. $\pm$ 26.&426. $\pm$ 36.&133. $\pm$ 57.&-&-61.4 $\pm$ 9.0\\
$[$Fe VII$]$$\lambda$6087& & & & & \\
\hspace{0.2cm}Broad&258. $\pm$ 30.&691. $\pm$ 74.&279. $\pm$ 69.&81. $\pm$ 30.&-73. $\pm$ 15.\\
\hspace{0.2cm}Narrow&-&-&-&-&-\\
$[$Fe X$]$$\lambda$6374& & & & & \\
\hspace{0.2cm}Broad&387. $\pm$ 52.&640. $\pm$ 80.&480 $\pm$ 180&-&-78. $\pm$ 28.\\
\hspace{0.2cm}Narrow&-&-&-&-&-\\
$[$Fe XI$]$$\lambda$7892& & & & & \\
\hspace{0.2cm}Broad&-&-&-&-&-64. $\pm$ 23.\\
\hspace{0.2cm}Narrow&-&-&-&-&-\\
$[$Fe XIV$]$$\lambda$5303& & & & & \\
\hspace{0.2cm}Broad&-&-&-&-&-\\
\hline
\end{tabular}

\end{table*}

\label{Balmer Line Results}
As expected, all our spectra are dominated by strong H$\alpha$ and H$\beta$ lines, which we show, together with their Gaussian fits, in Figs \ref{H_a} and \ref{H_b}, respectively.  We have been able to deconvolve these lines into their narrow and broad components for all the observed Seyfert 1 nuclei.  For six of the nine Seyfert 1s we observed the Balmer lines are best fitted with three components.  The three galaxies that do not fit this trend are NGC7469, VIIZW118 and NGC985, in which a third component does not provide a significant improvement to the fit.  In general, we find equivalent velocity width components (to within errors) in both the H$\alpha$ and H$\beta$ lines, giving us further confidence in the reality of each of the components.  The only case of which this is not strictly true is the intermediate component of 1H1934-063, although here the discrepancy is $<$20\%- typical of the size of the systematic errors on Balmer line components in other galaxies considered here.  

To assist in the discussion we have arranged our results in order of increasing H$\alpha$ intermediate component velocity width, which we find most closely matches the crude FWHM of the whole line.  Fitting the Balmer lines with Gaussians is complicated by the presence of a number of blended narrow and broad lines emitted by other species.  In particular, the H$\alpha$ line is blended with two often quite strong, narrow [N II] lines at $\lambda$6548 and $\lambda$6584.  Fortunately, as they are narrow these lines are easily fitted and removed when fitting the H$\alpha$ line.  More difficult to account for is the broad He I $\lambda$4922 line that forms the `red shelf' of the H$\beta$ line, and is discussed in detail by \cite{Veron02}.  If, during the initial fitting, there was evidence of a broad, redshifted component to the H$\beta$ line, we assumed that it was the result of this Helium line.  We then refit the blend, adding an additional Gaussian with its central wavelength fixed to 4922\AA (rest frame). This component is then excluded from further analysis of the H$\beta$ line.   A similar procedure was used to remove the less prominent red shelf on the H$\alpha$ line, caused by the presence of He I $\lambda$6678 emission.

\subsubsection{Balmer Ratios in the Velocity Components}
It is possible that emission from the BELR and NELR experience different amounts of reddening depending on the location of the dust within the emitting regions. It is also possible that the physical conditions  of opacity (self-absorption) and collisional effects will lead to differences in the Balmer ratios across the various velocity components of the Balmer lines. To test this possibility we have measured the H$\alpha$/H$\beta$ ratios for both emission components (see Table \ref{HaHbratios}).  We find that for all our targets the degree of reddening calculated from the Balmer decrement increases with decreasing component velocity widths (to within systematic errors), which could imply a higher concentration of dust in the NLR (narrow line region), as has previously been suggested by \cite{Netzer93}.

We note that in 1H1934-063 (and possibly Mrk618) the broad component has an H$\alpha$/H$\beta$ ratio that is less than the value predicted if there were no absorption whatsoever. Therefore it is clear that effects other than reddening are at play. However, we should be aware that the intensity of the broadest component is the most uncertain of all the fitted components as it is mainly determined by the weak wings on either side of the line centre.  These wings are often blended, especially in the case of H$\beta$, and in addition are highly dependent on the fitting of the continuum level.  In contrast, the intermediate and narrow components are better constrained by the profile of the line where the line flux is strongest.

Our work confirms that the overall profiles of H$\alpha$ and H$\beta$ are very often not the same. This implies differences in one or all the parameters of reddening, density and temperature, although this cannot be fully investigated without coverage of other Balmer lines (H$\gamma$, H$\delta$; e.g \citealt{Popovic03}, \citealt{LaMura07}).

\begin{table}
\centering
\caption{H$\alpha$/H$\beta$ ratios.  As we discuss in the text the systematic errors associated with the continuum flux level dominate over the random errors, so here we only give the systematic errors (in parentheses as percentages).  Many of the galaxies have different Balmer ratios for each velocity component, suggesting different amounts of absorption and/or physical conditions in each region.}
\label{HaHbratios}
\begin{tabular}{@{}lcccc@{}}
\hline
Galaxy&Total&Broad&Interm.&Narrow\\
\hline
Ark564&4.54 (15)&3.95 (5)&4.65 (20)&5.95 (25)\\
IZW1&4.54 (25)&3.70 (25)&6.23 (10)&4.95 (60)\\
1H1934-063&3.20 (20)&1.76 (10)&3.90 (25)&7.52 (15)\\
Mrk335&3.50 (5)&2.58 (5)&4.82 (5)&5.11 (10)\\
NGC7469&4.88 (10)&4.95 (10)&-&4.45 ($<$2.5)\\
Mrk618&3.31 (25)&2.41 (15)&6.20 (70)&6.57 (50)\\
IIZW136&3.83 (10)&3.37 (5)&3.75 (15)&5.92 ($<$2.5)\\
VIIZW118&4.13 (15)&2.97 (20)&-&4.94 (15)\\
NGC985&5.28 (10)&5.27 (10)&-&5.30 (15)\\
Mrk573&3.82 (10)&-&3.68 (5)&4.09 (15)\\
\hline
\end{tabular}

\end{table}

\subsubsection{Helium Emission}
Our spectra also cover a number He lines, of which He I$\lambda$5876 and He II $\lambda$4686 are two of the strongest.   Plots of these lines, including their Gaussian fits, are shown in Figs \ref{HeI} and \ref{HeII}, respectively.  We have been able to measure both these emission lines in seven of the nine observed Sy1 galaxies (the redshifts of IZW1 and IIZW136 place their He I$\lambda$5876 between the ranges of the Blue and Red/Short spectral settings).  Unfortunately, He II$\lambda$4686 suffers from strong blending with the Fe II lines and the H$\beta$ line just blueward and redward of it, respectively.  Despite our attempts to remove the blended Fe II emission there is still some remnants of this procedure (seen either in emission, or in absorption caused by overcompensation). There also appears to be an unidentified broad line present in the blue wing for the majority of the observed He II$\lambda$4686 that does not seem  to change after the Fe II removal.  Where this unidentified line is blended with the He II$\lambda$4686 line, we have fit it with a single Gaussian and have not included it in the interpretation of the He II line intensity and width.  Aware of these complications, we fit the He II line with multiple Gaussians where the addition of a second component provides a statistically better fit.

When it is present the broad component is generally redshifted with respect to the narrow component of the line, however, this may be an  artefact of line asymmetries caused by the Fe II removal. For the galaxies in which He II$\lambda$4686 is detected its strength relative to H$\beta$ varies considerably between galaxies, ranging from almost 25\% in Ark564, to less than 2\% in NGC985.

More accurately measured is the He I$\lambda$5876 line.  It is well detected in all our spectra that cover the region, and is easily deconvolved into its constituent velocity components.  In 4/9 of the Sy1-type AGN the broadest of these components is considerably broader than the broadest H$\beta$ velocity component, although the level of the continuum fit has a considerable effect on the width of the fitted broad component (see \S \ref{Discussion}).

\subsection{Narrow Emission Lines}

The narrow emission line region has been well studied in the past, and is therefore much better understood than the other  two non-spatially resolved regions discussed in this paper.  Because of this we do not provide a full analysis of the narrow emission lines. Instead we will simply mention the main points of the most prominent of these narrow lines.

\subsubsection{[O III] $\lambda\lambda$4959,5007}
All our spectra with the possible exception of IZW1, display strong [O III] $\lambda\lambda$4959,5007 lines, whose profiles all show asymmetries or have low intensity broad wings that cannot be modelled by a single Gaussian.  These asymmetries tend toward a blue wing, which when fitted using two Guassians, suggests a narrow (200km/s$<$FWHM$<$400km/s) component centred on the systemic wavelength, plus a broader component (350km/s$<$FWHM$<$1000km/s) that is generally blueshifted with respect to the narrow component.  This fact was first noted by \cite{Heckman81}, in which they attribute the effect to outflows with preferentially absorbed redshifted components, supported by the tight correlation between line asymmetry and H$\alpha$/H$\beta$ ratio.

\subsubsection{[S II] $\lambda$6717, 6732 doublet} 
As this feature is commonly used to determine the densities of the NLR of AGN, we chose to fit these lines with separate Gaussians, using multiple components if there was evidence of broad wings or inflections.  Comparing the [S II] 6716/6731 line ratio with published density relationships revealed no correlation between electron density and AGN type or width of H$\beta$.  This finding is consistent with the large scatter in the [SII] ratio vs. line width plot of \cite{Xu07} at low values of H$\beta$ FWHM. 

\subsubsection{Balmer line narrow component}
As a test to determine whether the narrow component of the Balmer emission is arising from the same region of the AGN as the  narrow forbidden line emission, we compare in detail the widths of the narrow components of the Balmer and [O III] lines.  Table \ref{width_sep_table} shows that 7/9 of the observed Seyfert 1s have narrow Balmer components that either lie in between or very close to the velocity widths of the narrow and broad components of the [O III] component widths.  The two galaxies in which this is not the case are IZW1 and VIIZW118.  As mentioned in the previous section, IZW1 has very broad, highly blueshifted [OIII] lines, so it is not surprising that the widths do not match in this case, and there is no evidence of a narrow component in the Balmer lines of VIIZW118. 

\subsection{Forbidden High Ionisation (Coronal) Lines (FHILs)}
\label{FHILS}
We detect the [Fe VII] $\lambda$6087 emission line in every galaxy.  The line profile in NGC7469 is best described by a two-Gaussian model.  Additionally, Ark564 and Mrk335 both appear to have a slight blue wing, but the improved fit of a 2 component model is not statistically significant.

We detect the [Fe X] emission line in seven of our spectra.  Our measurements are complicated by blending with the [O I]$\lambda$6363 line.  An estimate of the contribution from the [O I] line can be made by using the nearby [O I]$\lambda$6300 line as a calibration.  From atomic physics these two lines should have exactly the same profile, and the  intensities must have a ratio of 1:3. Using this information we are able to remove the contribution of [O I] from the [Fe X] blend. As with the He I and He II lines, measuring the broad component of the [Fe X] line is complicated by uncertainties in setting the local continuum levels and blends of multiple lines that form a pseudo-continuum. Despite these difficulties, we have measured the [Fe X] lines using Gaussian fits, which are shown in Fig. \ref{FeX}. Our analysis include a 2$^{nd}$ Gaussian component when statistically significant, although the majority of the spectra only require a single component.  All the [Fe X] lines we measure contain broad components that are broader than the [Fe VII] broad components. The two galaxies that show the narrowest Balmer lines also have the narrowest [Fe X] lines of the sample.  All the [Fe X] lines are blueshifted in the range between $\sim$150 and $\sim$700km/s, although there does not appear to be any correlation between blueshift and velocity width or relative intensity of the line between galaxies.

Unfortunately, about half of our spectra have rather poor S/N ratios in the region containing the [Fe XI] line.  This is partly attributed to telluric absorption (as discussed in \S\ref{ImRed}), and partly to interference fringes on the CCD chip around these wavelengths.  Due to this, we only have measurements of this line for three galaxies, Ark564, 1H1934-063 and Mrk335, as shown in Fig. \ref{FeX}.  The widths of the [Fe XI] lines reveals that they are similar to the [Fe X] line velocity widths in Ark564 and 1H1934-063.  Mrk335 seems to have a wider broad component in [Fe X], but this value is uncertain because of poor continuum determination.

The highest ionisation line species we observe, [Fe XIV]$\lambda$5303, lies within a region that is heavily affected by the presence of  Fe II emission.  Despite our attempts at mitigating the effects of Fe II emission, we have still had great difficulty in isolating the [Fe XIV]$\lambda$5303 line.  The [Fe XIV] line is also strongly blended with the [Ca V]$\lambda$5309 line.  Although we cannot measure the [Fe XIV]  line accurately enough to give robust quantitative results, we are confident of its presence in two of our sample, Ark564 and 1H1934-063 ($\Delta\chi^{2}>$500 and $>$1000, respectively, for 1 added Gaussian- corresponding to a $\gg$10$\sigma$ detection in both targets).  In both cases we are able to separate the feature into 2 Gaussian components ($\Delta\chi^{2}$=52 and =85 between 1- and 2- component fit, again, corresponding to a $>$10$\sigma$ detection of two separate lines in both targets), which we interpret as [Fe XIV] and [Ca V].  We take further confidence in this from the fact that the velocity shift of the [Fe XIV] line is similar to that of the [Fe X] and [Fe XI] lines.  Differentiation of the [Fe XIV] and [Ca V] lines is helped by the fact that the former is more strongly blueshifted thereby increasing the observed separation from the intrinsic 6\AA\ to 15-20\AA.  That the ionisation potential of [Ca V] is significantly less than [Fe XIV] (67eV and 361eV, respectively) provides further evidence that blueshift increases with the ionisation potential of the emitting species. 

\subsection{Mrk 573}
Mrk573 is a relatively well studied nearby ($z=0.0170$) Seyfert 2 galaxy, which has been shown to contain a hidden BLR that is detectable in polarised light (\citealt{Nagao04}) suggesting the presence of obscuring matter along our line of sight to the BLR, possibly the putative torus. We therefore observed this galaxy as a comparison for the Seyfert 1s, but without the complication of broad emission components on both the permitted lines and the FHILs.

As \cite{Erkens97} had already found, this Seyfert 2 contains strong FHILs. We chose to observe this galaxy at higher spectral resolution and S/N levels, to determine if these lines had any structure in their profiles that might indicate outflows or turbulence in the FHIL emitting regions of Seyfert 2s. 

The various line profiles extracted from the spectrum of Mrk573, with their associated Gaussian fits are shown in the last panel of the all the figures discussed previously.   

\subsubsection{Narrow Emission Lines}
As expected for a Seyfert 2 spectrum all the emission lines in  Mrk573 are considerably narrower than the permitted lines observed in Seyfert 1 galaxies.  However the strong [O III] and Balmer emission lines show signs of symmetric, broad wings and were modelled with multiple Gaussian components, while all the weaker lines were satisfactorily fit with a single component.  Excluding the broad wings of the strongest lines, all the emission lines had similar velocity widths of FWHM  $\sim$300km/s.  We also detect strong He II$\lambda$4686 emission in this galaxy.

\subsubsection{Forbidden High Ionisation Lines}
Of note is the large number of FHILs that we detect in Mkn573.  They are comparable to the most highly ionised Seyfert 1 spectrum in our study, Ark564.  In addition to [Fe VII]$\lambda$6087 both [Fe VII]$\lambda$5159 and [Fe VII]$\lambda$5722 are well detected.  In contrast to the trend found in the Seyfert 1s, the highest ionisation lines in Mkn573 show no sign of asymmetry or wing broadening, and have thus been fitted with a single Gaussian.  The resulting profiles of [Fe X] and [Fe XI] have widths and velocity shifts similar to that of [Fe VII].  We note that these lines appear to be redshifted relative to [S II], but are consistent with the velocity of the Mgb absorption lines.  We do not detect [Fe XIV]$\lambda$5303 in this object.

\section{Discussion}
\label{Discussion}
\begin{figure*}
\begin{center}
	\includegraphics[width=18cm, height =
	13cm]{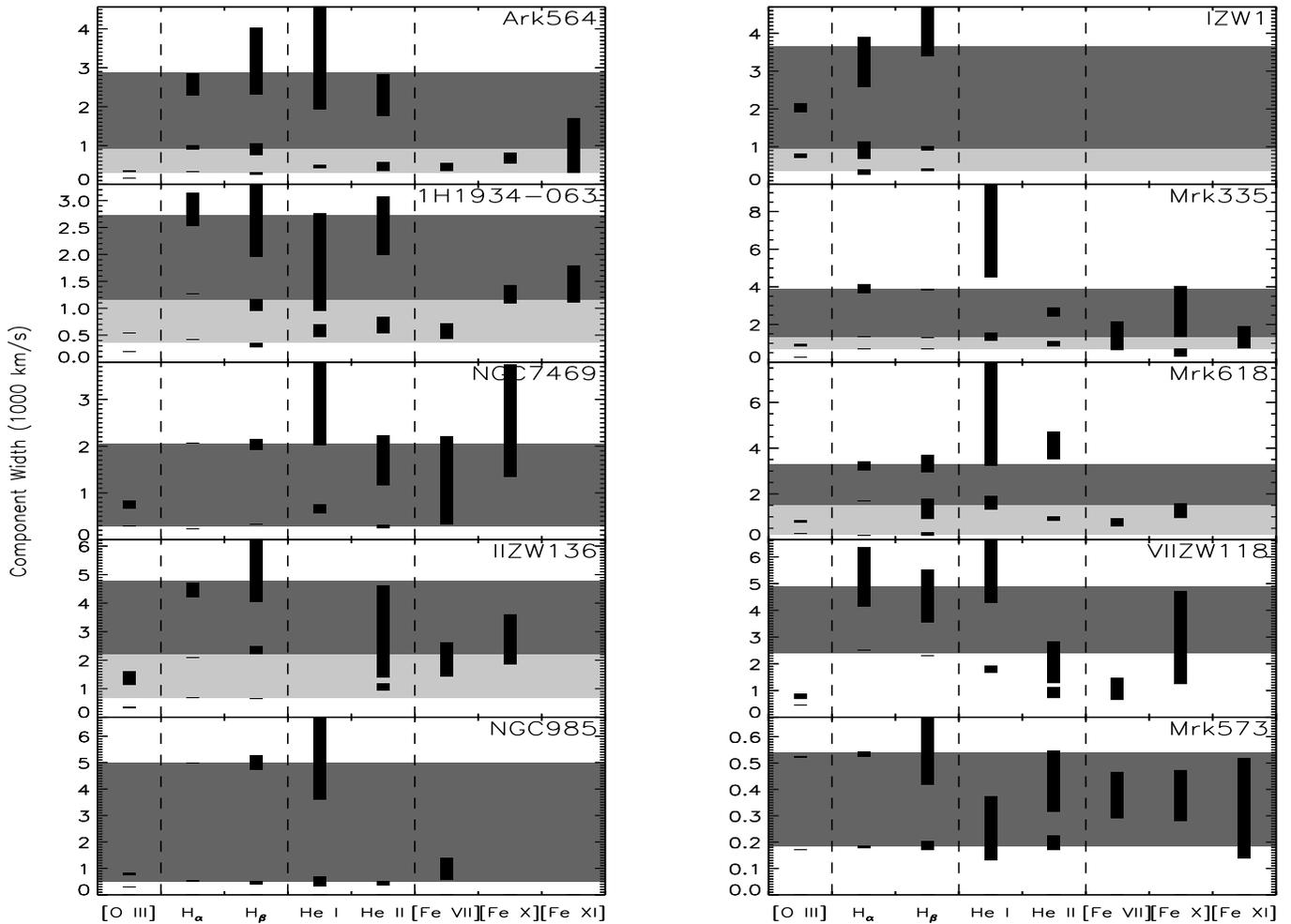}
\end{center}
\caption{The component widths of each of the lines discussed in detail in this paper.  The vertical height of each block represents the extent of both the systematic and random errors associated with each line model  component.  The shaded areas represent the regions between the `narrow' and `intermediate' and `intermediate' and `broad' widths, averaged between the H$\alpha$ and H$\beta$ components (only one shaded region is shown in cases where the permitted line fit required only 2 components).  This clearly shows that the FHILs tend to have widths between those of the traditional broad and narrow emission lines.}
\label{Blocks}
\end{figure*}

\begin{figure*}
\begin{center}
	\includegraphics[width=18cm, height =
	13cm]{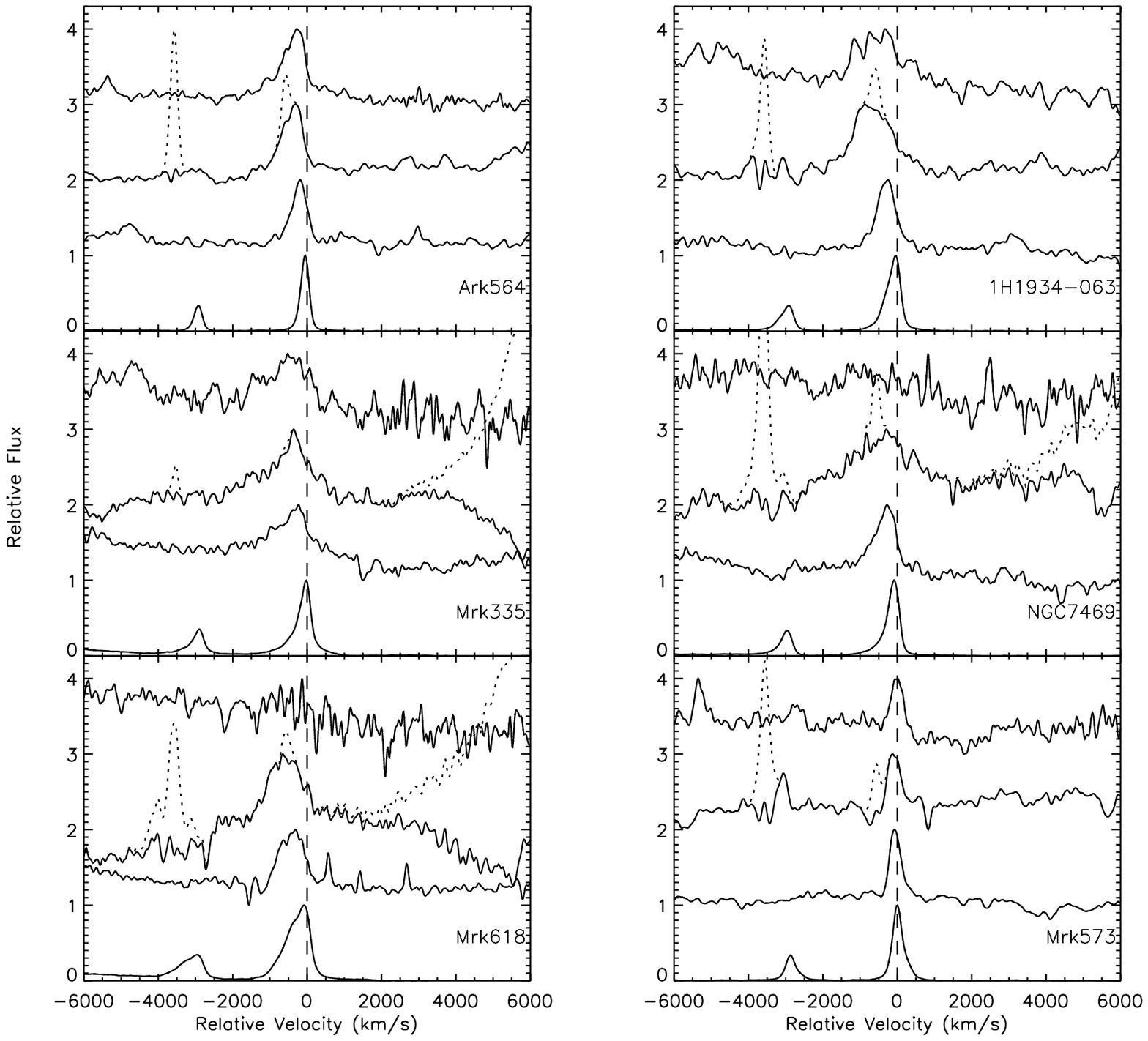}
\end{center}
\caption{Comparing the profiles and velocity shifts of the high ionisation lines.  We include the narrow [O III] line as a comparison.  Profiles shown are (from bottom) $[$O III$]$$\lambda$5007, $[$Fe VII$]$$\lambda$6087, $[$Fe X$]$$\lambda$6374 and $[$Fe XI$]$$\lambda$7892.  The dotted lines indicate removed features, such as the $[$O I$]$$\lambda$6363 line that is often blended with the $[$Fe X$]$$\lambda$6374 line and the blue wing of the H$\alpha$ line.  It is generally seen that line widths and shifts increase with ionisation potential.}
\label{comparison}
\end{figure*}

As described in detail in the previous sections, we have been able to decompose many of the optical emission lines into multiple components that span a range in velocity widths and velocity shifts- indicating that the spectra of Seyfert galaxies cannot be considered simply in terms of narrow and broad emission lines.  This is clearly shown in Fig. \ref{Blocks}, in which we display the range of the component widths for each line species which  we will discuss in detail.

While traditionally the permitted lines of Seyfert galaxies were considered to be a blend of emission from both the broad and narrow line regions, we find evidence of a very broad (2500km/s $\lesssim$ FWHM $\lesssim$ 7000km/s) third component in 6/9 of the Seyfert 1s  (all six of them NLS1s) we observed.  This extra component was previously noted in the NLS1 galaxy KUG1031+391 by \cite{Mason96} with a measured FWHM ($\sim$2500km/s) more typical of the broad lines in a Seyfert 1.  It was suggested by \cite{Goncalves99} that the broad wings of the Balmer lines in KUG1031+391 were better fit with a single Lorentzian profile, although \cite{Dietrich05} have since shown that this function's profile overestimates the core of the line, while underestimating the flux in the wings, in their study of 12 NLS1 galaxies. In our appendix, we demonstrate a 3 Gaussian component fit is superior to a Lorentzian-plus-Gaussian fit.  These models reveal a very broad component in a further 6 NLS1 galaxies, in which we measure broad components of similar widths when fitting both the H$\alpha$ and H$\beta$ lines independently.  We note that care must be taken when interpreting lines with multiple components.  It is unlikely that the dense emitting gas around the central engine can be simply divided into discrete velocity-width bins (in this case, three). The true situation is that we see an  amalgam of emission from a ensemble of clouds along our particular line of sight with a wide range of velocity widths (see, for example, \citealt{Osterbrock93b}).  

As discussed by \cite{Dietrich05}, the fact that we see emission line components with widths similar to those of the broad emission lines of BLS1 galaxies indicates that emission from this BLR may be reduced in NLS1s, rather than absent altogether.  This could imply either:

\begin{itemize} 
\item that we should no longer assume that black hole masses in NLS1s are smaller than in BLS1s.  This  may appear to contradict other evidence of reduced black hole masses in NLS1s, including values derived from stellar velocity dispersion ($\sigma$) and X-ray variability measurements.  It has been claimed, however, that the validity of the $M_{BH}$-$\sigma$ relation shown in Figure 2 of \cite{Barth05} is uncertain at low values of values of $\sigma$.  This is particularly evident when one considers that the $M_{BH}$ estimate for M33 is a hard upper limit (and more likely that this galaxy does not contain a central black hole; \citealt{Gebhardt01}) and the mass of the black hole in NGC4395 has been revised upwards by an order of magnitude (\citealt{Peterson05}).  Taking this into account, it is possible that the $M_{BH}$-$\sigma$ relationship shows an upturn at low values of $\sigma$, strengthening the claim that we may expect broader components in the Balmer lines of NLS1s.  Also, although detailed long term X-ray variability measurements can be used to derive black hole masses using their power density spectra (see e.g. \cite{McHardy06}), isolated short term X-ray variability may not be a reliable indicator of the black hole mass.  This has recently been shown by \cite{Reeves02}.
\item that NLS1 galaxies \textit{do} contain lower mass black holes, but the radius of the broad line region scales in proportion to the mass of the black hole.  This would then mean that other factors determine the observed widths of the lines. 
\end{itemize}

For both these explanations, we would need a theory to explain the reduced relative luminosity of the very broad component.  This could be disruption by the inner parts of an energetic wind originating from within the accretion disc that would have the largest effect on a nearby broad line region, or the strong UV/soft X-ray flux seen in the SED of many NLS1s in the form of the soft-excess, or a mixture of both.

We find that the  permitted Helium line profiles differ considerably from the Balmer lines for the sub-set of galaxies (6/7) in which the Helium line can be measured.  This fact has been reported by \cite{Shuder82}, and has since been briefly mentioned in the course a number of permitted line studies (e.g. \citealt{Almog89}, \citealt{Dietrich05}), in which the broad component of the He I $\lambda$5876 line was found to be broader than the corresponding Balmer line component.  We too find that this is the case in three of the galaxies in which this line was observed.  This suggests some form of segregation of the broad line region, in which different species of permitted lines are emitted preferentially in regions with differing conditions and  kinematics.  This is also suggested by the results of the locally, optimally emitting cloud (LOC) models of \cite{Baldwin95}.  Because we are uncertain of the processes that cause  broadening of the permitted lines we cannot rule out diverse explanations, for example, that a greater proportion of the He I emitting region lies closer to the central engine than the Balmer line emitting region, because of the higher ionisation of Helium compared with  Hydrogen, or alternatively whether the Helium clouds are undergoing greater turbulent motion or are outflowing/inflowing at higher speeds as has been discussed by e.g. \cite{Collin-Souffrin88}.  However, caution must be taken when interpreting the widths of the He lines as we have shown that the level of the underlying continuum can have a considerable effect on the properties of the fitted components.  

Due to the difficulties with blending, we have been unable to measure the broad component of the He II$\lambda$4686 line to the same degree of accuracy as He I and Balmer lines. Despite this, there is some evidence that it has a broad component with a FWHM which is significantly less than the corresponding He I and Balmer line components (see, in particular, Ark564, Mrk335, NGC7469 and VIIZW118 in Fig. \ref{Blocks}).  If this is the case, and is not the result of measurement errors or Fe II removal, then this runs counter to the prediction from photoionisation models of \cite{Korista04}- that the broad components of the permitted lines should increase in width with increasing ionisation potential.  \cite{Landt07} find a similar situation when comparing the widths of the permitted lines in both optical and near-infrared spectra.  They speculate that this effect may be caused by an accelerating outflow, in which the He$^{+}$ gas is closer to the source of the ionising continuum than He$^{0}$, but is outflowing more slowly as it has yet to be accelerated to its terminal velocity.  This would be a very important result, but clearly much more work is necessary to determine the true kinematics of the He I, He II and the Hydrogen emitting clouds.

Evidence of emission components with widths between the conventional broad and narrow lines is also seen in the forbidden lines of all the objects we observe, in the form of a broad blue wing.  This is a well known feature that has been studied in depth since its first detection and description by \cite{Heckman81}, in which they attribute the blue wings of the [O III] lines to outflowing gas, which is likely to have a density similar to the traditional NLR (i.e. $n_{e}\sim$10$^{4}$cm$^{-3}$), interspersed with dust that preferentially obscures the gas outflowing in a direction away from us.  More recent studies (\citealt{Whittle92}, \citealt{Christopoulou97}, \citealt{Veilleux01}, \citealt{Veron01}) confirm that the [O III] emission is likely to arise from an ambient region at the same redshift and rotational velocity as the host galaxy, and an `outflowing' region that is distributed in two wide angle ionisation cones.  We note that  the critical densities of [O III]$\lambda$5007, [Fe VII]$\lambda$6087 and [Fe X]$\lambda$6374 (log($n_{e}$)$\sim$5.8, $\sim$7.6, $\sim$9.7, respectively; \citealt{Appenzeller88}) suggests that the outflowing gas responsible for [O III]  blue wings is likely to have a lower density than the FHIL outflows.  

Strong evidence of gas with kinematics between the classic narrow and broad emitting regions is found in the form of the FHILs.  As was noted in \cite{Erkens97} and \cite{Rodriguez06}, we find that as well as being significantly broader than the lower ionisation forbidden lines, these lines show increasing blueshift with increasing ionisation potential.  This is clearly shown in Fig. \ref{comparison}, in which we plot the line profiles in terms of velocity shift from systemic for the [O III]$\lambda$5007 and the Fe FHILs, which suggests an outflow that is possibly related to that responsible for the broad, blue wing of the [O III] lines.
 
We note that, where it is measured, the width of the [FeVII]$\lambda$6087 line is generally between that of the narrow and intermediate Balmer line components.  This, combined with findings that this line also tends to be blueshifted with respect to the permitted and low ionisation forbidden lines suggest this line is emitted from a  kinematically distinct region.  We find that the relative intensity of the [Fe VII]$\lambda$6087 is reasonably constant among our small sample, at 2-4\% flux of the H$\beta$ line in all the galaxies we observed.  A larger, more homogenous sample would be required to fully determine correlations between the [Fe VII] line properties and AGN classes, such as NLS1s and BLS1s. 

Our fits to the [Fe X]$\lambda$6374 line suggest that this species is formed in yet another kinematically distinct region, as was noted by \cite{Rodriguez06} in nearby Seyfert 1 and Seyfert 2 galaxies.  Although approximately twice as strong as [Fe VII]$\lambda$6087 in all the galaxies in which it is detected, this line is more poorly defined because of its much broader profile and its blending with other emission lines.  In many of the galaxies it has a broad component that approaches the FWHM of typical of broad permitted lines, and shows little or no evidence of a narrow component, possibly implying that the Fe$^{9+}$ ion is only created in regions near to the central source of ionisation, which would agree with the predictions of the photoionisation models of \cite{Ferguson97}.  As found for [Fe VII] the relative intensity of the [Fe X] line appears to be similar in all the galaxies we observed ($\sim$8\% H$\beta$, or 2$\times$[Fe VII]$\lambda$6087).  We note that the almost constant value of [Fe X]$\lambda$6374/[Fe VII]$\lambda$6087$\sim$2 appears at odds with the findings of \cite{Nagao00}, who generally found lower values for this ratio in a larger sample of AGN (including NLS1s, BLS1s and Sy2s).  We note, however, that the systematic errors associated with the fitted continuum level has a significant effect on the measured intensity of the broad components of these weak lines, which could account for these differences.

In contrast to generally similar [Fe X]$\lambda$6374 relative line intensities among our sample, we find that the [Fe XI]/[Fe X] ratio covers a wide range e.g. from $\sim$1 in 1H1934-063 to $<$ 0.03 in NGC7469 (taking 90\% upper confidence limits for the [Fe XI] line).  This is despite the fact that these two species have similar ionisation potentials (235eV and 262eV for [Fe X] and [Fe XI], respectively).  The origin of this wide range of ratios is not well understood.  However, it is in agreement with the photoionisation models of \cite{Ferguson97} who point out that the [Fe XI]$\lambda$7892 line is highly sensitive to the density and ionisation parameter of the FHIL emitting region.  In addition, if we assume that [Fe XI] is produced by photoionisation (as suggested by \citealt{Penston84}) it is possible that the range in the line ratio is also a function of the continuum shape in this energy region.

Despite the difficulties in extracting the [Fe XIV]$\lambda$5303 mentioned in \S \ref{FHILS}, we are able to qualitatively confirm the presence of this line in two of our sample, Ark564 and 1H1934-063.  These galaxies also display strong [Fe X]$\lambda$6374 and [Fe XI]$\lambda$7892 lines, and we are able to show that the blueshift of the [Fe XIV]$\lambda$5303 line is similar to the blueshift of these other lines, giving us further confidence in this detection.  However, it should be emphasised that the systematic errors associated with this measurement caused by the removal of the FeII blend is unknown and are likely to be large, we therefore make no attempt to interpret the intensity or width of this line.

Finally, turning to the case of the only Seyfert 2 in our sample, it is clear from Fig. \ref{Blocks} that the highest ionisation lines in Mrk573 trace a kinematically different region than both the permitted and other forbidden lines. Apart from a weak, broad component of the [Fe VII]$\lambda$6087 line, the widths of FHILs are all relatively low (FWHM $\sim$ 400km/s).  These results agree with those of \cite{Prieto05} and \cite{Rodriguez06}, who found that there is significant [Si VII] (a FHIL observed in the near-IR) emission in regions up to $\sim$150pc from the nuclei in four nearby Seyfert 2 galaxies. However, it is difficult to reconcile their results with our findings for BLS1s and NLS1s in terms of the standard AGN model.  Strong [Fe XI]$\lambda$7892 is detected in all four Seyfert 2 galaxies observed by \cite{Rodriguez06} and in our work. However, this line is detected in less than half of the Seyfert 1s that have been observed by us and in \cite{Erkens97}. If, using the standard model we expect that the NLRs of both Seyfert 1 and Seyfert 2 galaxies are essentially the same then we would expect from the Seyfert 2 results to detect a narrow  [Fe XI]$\lambda$7892 line in a much larger proportion of Seyfert 1 galaxies. Of course, sample selection effects may be important and it will be necessary to observe larger samples of the various classes of Seyferts at comparable signal to noise to make further progress in this area.  

\section{Summary}

The main conclusions from our study can be summarised as follows;

\begin{itemize}
\item The emission lines in the spectra of both BLS1s and NLS1s appear to be emitted from more zones than the traditional narrow and broad line regions.  This is particularly evident in the profiles of the FHILs, but also in the profiles of the broad, permitted lines of H$\alpha$ and H$\beta$.  The lines from the FHIL region typically have a FWHM of 1000 - 3000km/s, and are significantly narrower than the broadest components of the permitted lines observed in the same object. 

\item All of the NLS1s in our sample show evidence of a very broad component in their permitted lines.  This component has a FWHM that is more typical of that observed in the broad lines of BLS1s, i.e. $\sim$3000km/s.   Although the FWHM measurements are a product of our line modelling process, the point that the broad wings of the permitted lines extend to up to 3000km/s in even the `narrowest' NLS1s indicates that there exists emitting clouds with much higher relative velocities than has typically been assumed in this sub-set of Seyferts. However, this component appears to be greatly suppressed in the case of NLS1s. We speculate this suppression might be either a result of the destruction of the standard BLR by an energetic wind or by the increased high energy photon flux of NLS1s, or both.  We propose two methods of production of this extra broad component: 1) a higher mass black hole than is expected for NLS1s and a BLR radius more typical of BLS1s, or 2) a low mass black hole and an equivalently smaller BLR radius.  

\item In a number of our sample the He I $\lambda$5876 line contains a component that is broader than the broadest H$\alpha$ and H$\beta$ components.  This may be an indication that the permitted He lines are preferentially emitted closer to the central engine than the H permitted lines, or that a different mechanism is causing the broadening of these emission lines. 

\item We detect [Fe XI]$\lambda$7892 in a few cases where the data has sufficiently high signal to noise.  Despite [Fe X] and [Fe XI] having similar ionisation potentials, the [Fe X]/[Fe XI] ratio covers a wide range within our sample.  It is therefore possible that this ratio is very sensitive to the physical conditions of the emitting region as well as the shape of the ionising continuum.  This relationship will be investigated in a follow up study.

\item Strong, narrow FHILs are also detected in the Seyfert 2  Mrk573 (as also noted by \citealt{Rodriguez06}).  Since it has been shown that this AGN contains a hidden BLR, this result indicates that FHILs can be formed in regions beyond the extent of the putative dusty molecular torus.

\end{itemize}

We would like to express our gratitude to Dr. N. Schurch and Dr. J. Gelbord for the considerable help they provided on the line modelling processes, to Dr. D. Alexander for the comments he provided throughout this work, and to Dr. T. Roberts for his expertise and assistance during the observing run.  We would also like to thank the anonymous referee for their helpful comments.  J. R. Mullaney is funded by a STFC PhD research grant. 

\bibliography{../Bibliography/Bibliography}
\bibliographystyle{./TeXStyles/mn2e}
\bsp

\appendix
\section{Multiple component fits to the permitted emission lines}
A statistical goodness-of-fit test is required in order to conclusively determine whether multiple Gaussians or a Lorentzian plus Gaussian model best represent the shape of the observed broad permitted emission line profiles.
 
In the literature the principal statistical test for this is the $ \chi^{2}$ fit statistic, which takes account of both the data points and their associated errors.  When we apply this statistic to our Balmer line data we find that the $\chi^2$ values of the three-Gaussian (3G) models are significantly lower than those of the Lorentzian-plus-1-Gaussian (L+1G) models.  For example, in the case of the H$\beta$ line in Ark564, the $\chi^{2}$ statistic for the 2 Gaussian is 6077.97, L+1G: 2249.38 and 3G: 2186.66.  The best-fitting L+1G and 3G models both provide a much better description of the data than the two-Gaussian model.  Comparing these two models, we find that the 3G model improves the fit by $\Delta\chi^2 = 62.72$ with the addition of just three additional free parameters.  The probability of three extra parameters improving the $\chi^2$ by as much as 21.10 purely by chance is 0.01\%.  The $\Delta\chi^2$ is at least this large for each line in tables 2-4 that is described with three components.  Thus, we assert that the 3G model provides a better fit with $>$99.99\% confidence.

We emphasise that the \textit{reduced} $\chi^2$ of all of the above models is $\gg1$, so none of these provide a formally good fit to the data.  This is an inherent problem when fitting simple models to large-scale features in high signal-to-noise data.  The random error associated with the flux in each spectral bin is small  and thus the potential contribution of any individual bin to the  total $\chi^2$ is large.  Small `bumps-and-wiggles' in the data that  are not reproduced in the model will thus contribute strongly to the  total $\chi^2$.  Although each of these deviations may well be a real  feature of the source, we are not presently interested in complicated  models that could reproduce such details.  The inevitable consequence  of leaving such features unmodelled is that the reduced $\chi^2$ of  the best-fitting model will invariably be larger than 1.  In this paper we are only interested in the large-scale structure of  the lines.  If we want our relatively simple models to fit the  spectra well we need to minimise the impact of the small-scale bumps-and-wiggles.  This can be accomplished by smoothing the data.  So  long as the smoothing radius is not too large, the gross features in  which we are interested will be preserved.

To further test the L+1G and the 3G models we have applied both to a  sequence of increasingly smoothed spectra.  We convolve each spectrum  with a range of Gaussian kernels, varying the Gaussian FWHM from 1.0\AA\ to 25.0\AA\ in 0.5\AA\ steps.  To the lines in each we fit the L+1G and 3G models and measure the resulting reduced $\chi^2$.  We note that by using the reduced $\chi^2$ we take into account  the greater number of free parameters (and therefore greater  flexibility) associated with the 3G model.  As the smoothing radius increases,
(1) the small-scale structures in the data get washed out,
(2) the information content of the spectrum decreases, and
(3) the reduced $\chi^2$ of the best-fitting large-scale line models decreases.

With the appropriate amount of smoothing, the reduced $\chi^2$ of a model will approach 1.0 and the model can then be said to provide a good  description of the (remaining) data.  Any additional smoothing will decrease the reduced $\chi^2$ below 1.  When this happens the model is no longer appropriate because the spectrum has lost so much information that it no longer requires a  model of this complexity.  In effect, the model has too many degrees of freedom for such a  simplified spectrum and its use over-interprets what is left of the data.  We find that for the lines measured, the 3G models require less  smoothing than the L+1G models to reach a reduced $\chi^2$ of 1.  We therefore consider the 3G model to be the more appropriate model because it describes the spectra well when less of the original  information content has been lost.

\begin{figure}
\begin{center}
	\includegraphics[width=8cm, height = 5.0cm]{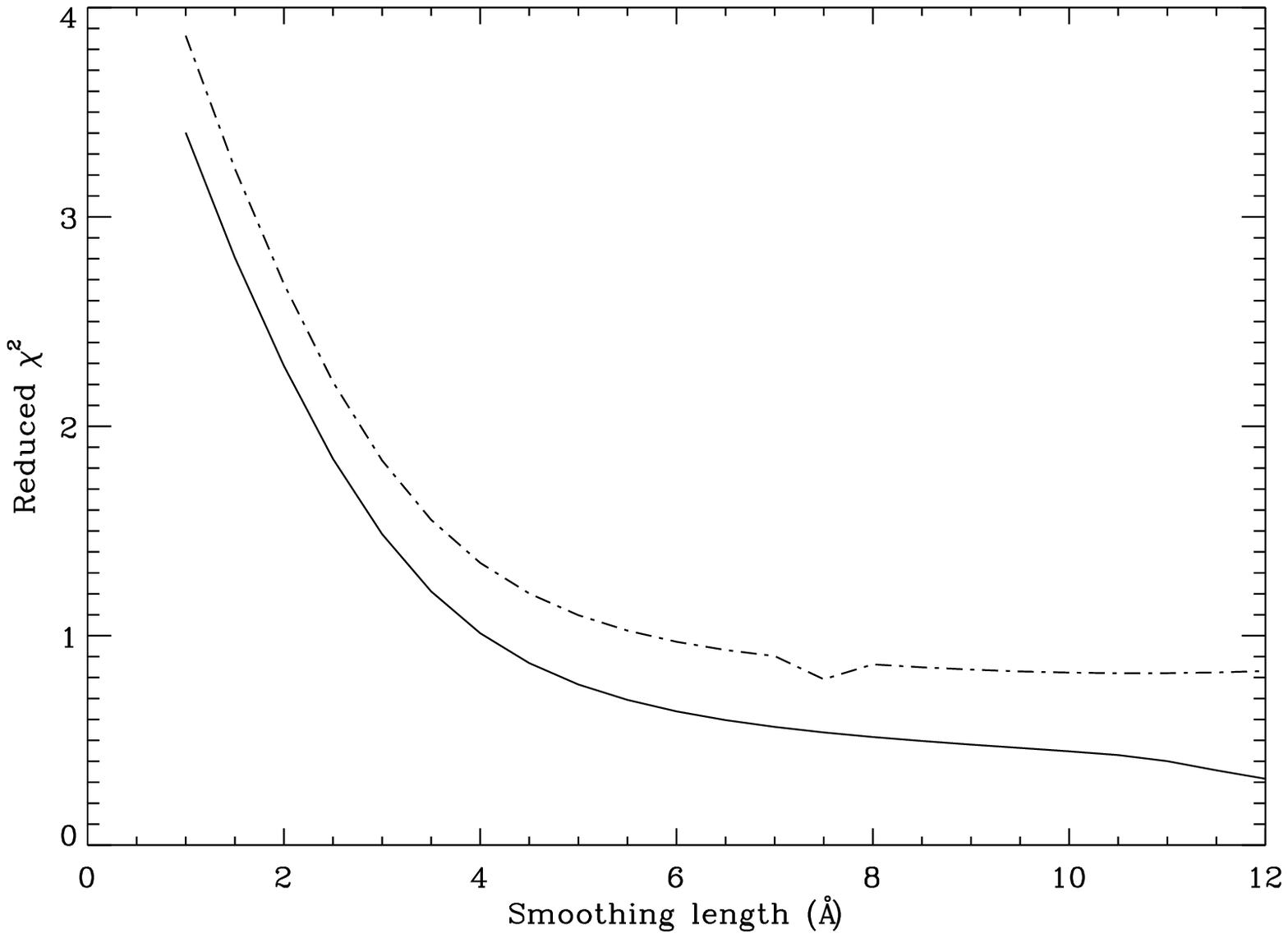}
\end{center}
\caption{The effect of smoothing on the reduced-$\chi^{2}$ statistic for the Ark564 H$\beta$ line (here, smoothing length refers to the FWHM of the Gaussian smoothing kernel).  The dotted line corresponds to the L+1G fit, and the solid line refers to the 3G fit.  It is evident that, irrespective of the smoothing length used, the 3G fit is statistically better than the L+1G fit.  That the reduced-$\chi^{2}$=1.0 point is reached by both fitting methods at smoothing lengths significantly below the maximum smoothing length revealed in \ref{maxsmooth} confirms this is still within the regime at which the narrow component of the H$\beta$ remains in tact after smoothing}
\label{chi_vs_smooth}
\end{figure}

As an example, fig. \ref{chi_vs_smooth} shows a comparison of the reduced-$\chi^2$  values for both the 3G and L+1G models applied to the H$\beta$ profile  of Ark564 as a function of smoothing width.  The reduced-$\chi^2$ of  the 3G model reaches 1 with a smoothing kernel of $\sim$4.0\AA\ FWHM  whereas that of the L+1G model requires a kernel of $\sim$5.6\AA\  FWHM.  Thus the 3G model reflects more of the information contained  in the original spectrum.  Quantitatively, when the 3G model fitted to the smoothed spectrum has  a reduced-$\chi^2$ of 1.00 the reduced-$\chi^2$ of the L+1G model is 1.35, indicating that the former is a better representation of the  data with $>$99\% confidence.  Moreover, the fact that the reduced-$\chi^2$ curve of the 3G model is always below that of the L+1G model indicates that the 3G model always  provides a better representation of the data regardless of the level  of smoothing applied.

\begin{figure}
\begin{center}
	\includegraphics[width=8cm, height = 5.0cm]{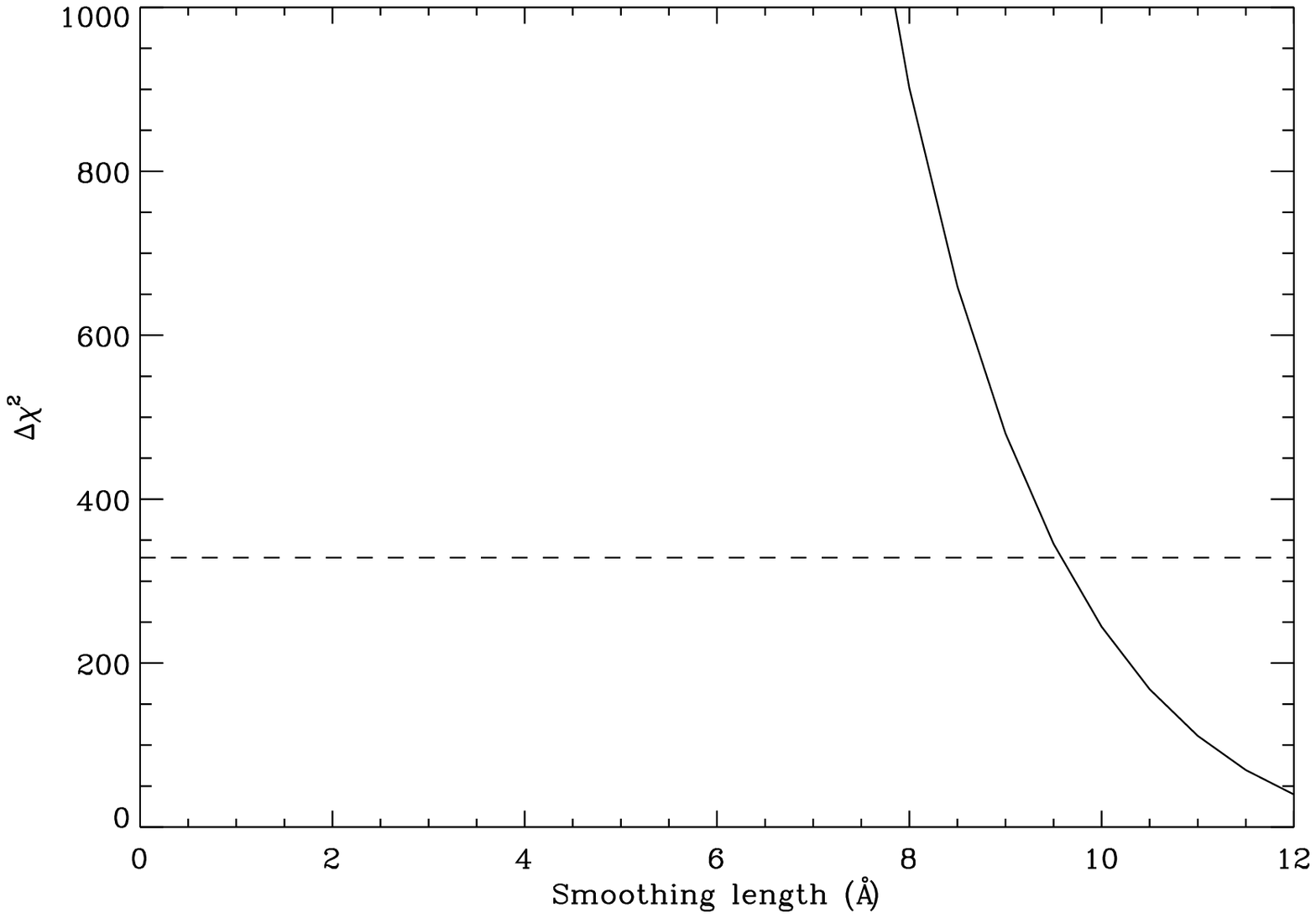}
\end{center}
\caption{Change in $\chi^{2}$ after the removal of the narrow component of the H$\beta$ line of Ark564 when fitted with a 3G model.  The dotted line shows the 90\% confidence level that the narrow component is detectable when performing the fit.  Using smoothing lengths significantly less than point at which the two lines intersect effectively ensures that we are smoothing out the small deviations discussed in the text, but retaining the overall profile of the H$\beta$ line. }
\label{maxsmooth}
\end{figure}

It is important to note, however, that care must be taken to choose an appropriate level of smoothing; i.e. smoothing such that the weak, narrow, `bumps-and-wiggles' are removed from the data without strongly affecting the broad line profile itself. Our primary concern is to ensure that the smoothing does not significantly alter the narrow component that makes up the core of the broad Balmer line profile. To ensure that we were still able to measure the properties of the narrowest component after smoothing, we fit a 3G model to the unsmoothed data, recording the properties of the narrowest component. Again, we then convolve the line profile with a Gaussian kernel whose width increases from 1.0\AA ~FWHM to 25.0\AA ~FWHM, in steps of 0.5\AA. At each step the smoothed data are re-fit with a 3G model in which the width of the narrowest component was fixed at the width of the narrowest component, based on the fit to the unsmoothed data. After calculating the $\chi^{2}$ we then subtracted the narrowest Gaussian from the model. The change in $\chi^{2}$ that results from the removal of the narrowest component of the model reveals how well the narrow component is detected for a given level of smoothing, and so places an upper limit on the appropriate smoothing value for an individual line profile depending on how stringently we wish the detection of the narrowest component to be. Fig. \ref{maxsmooth} shows the change in $\chi^{2}$ when the narrowest component is removed, as a function of smoothing width, in the case of Ark564. For each emission line we limit our chosen smoothing scale to be much less, that is by at least a factor of ~2, than the maximum smoothing scale at which the narrow component can still be detected at the 90\% confidence level. This is to ensure that the narrowest component of the line is not significantly affected by the smoothing process. This places an upper limit on the appropriate smoothing width of $\sim$4.75\AA. We note that the `best-fit' smoothing width for Ark564 is $\sim$4\AA, which is comfortably below this upper limit.

\begin{figure*}
\caption{Individual line plots and component fits.  For this and all the following plots of the line profiles, we only show the spectra for the galaxies in which the line is detected and measured.  In each panel, the vertical dashed lines show the rest wavelength of the line of interest.  Dotted lines reveal where we have removed a blended line to assist component fitting.   Colour coding- Red: Full line model, Magenta, Green, Blue: Individual lines components (broadest to narrowest), Orange: Blended component not associated with measured line (e.g. [N II] blended with H$\alpha$).  For the case of H$\alpha$, shown here, we have removed the [N II] lines in Mrk573, as described in the text.  }

\label{H_a}
\begin{center}
	\includegraphics[width=18cm, height = 15cm]{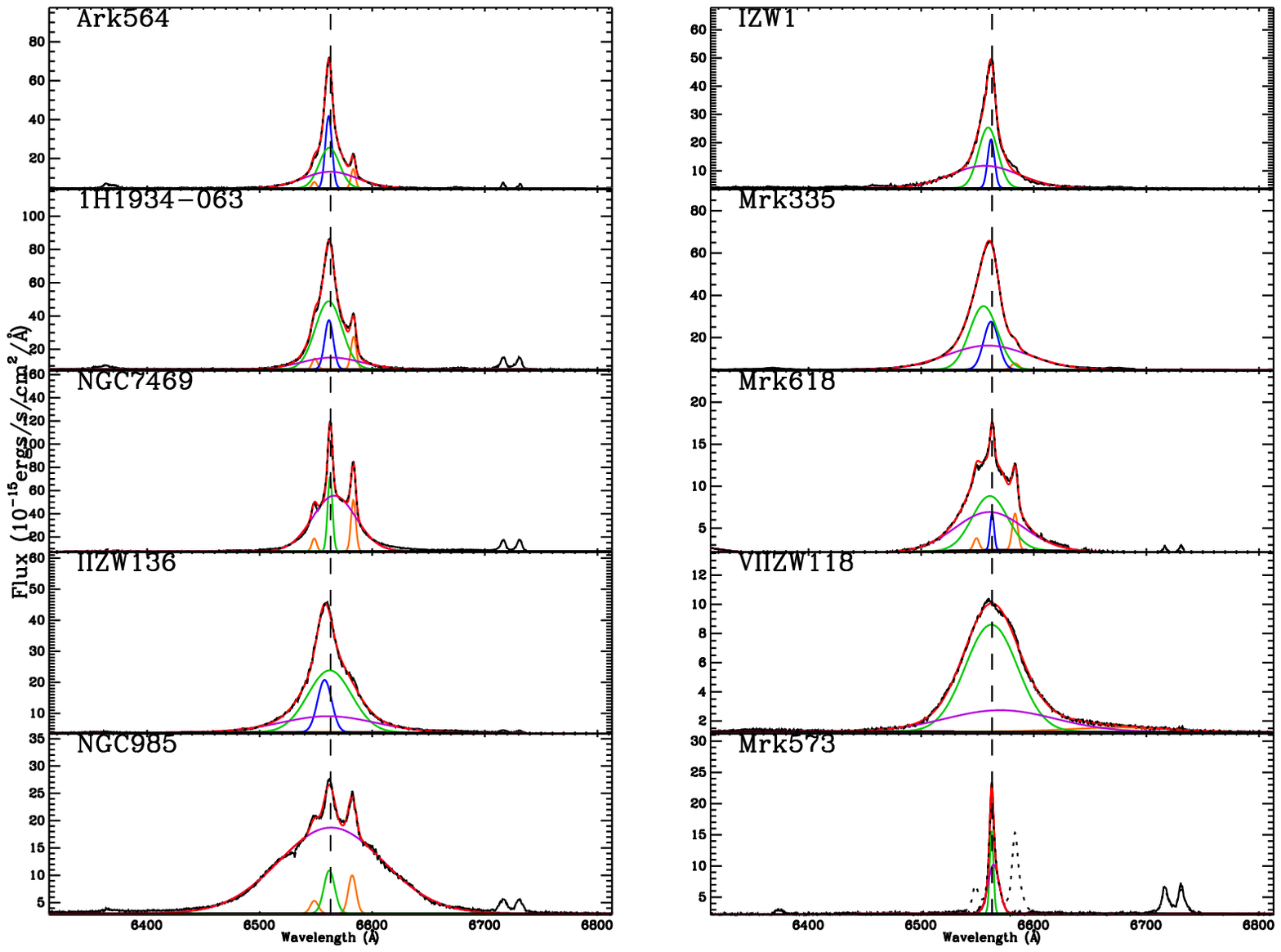}
\end{center}
\end{figure*}

\begin{figure*}
\caption{H$\beta$ \& [O III]- Dotted lines show where we have removed the [O III]$\lambda$4959 line in order to more accurately fit the H$\beta$ lines.  In the case of VIIZW118, we have also removed the He I$\lambda$4922 that forms a strong red shelf in this spectrum.}
\label{H_b}
\begin{center}
	\includegraphics[width=18cm, height = 15cm]{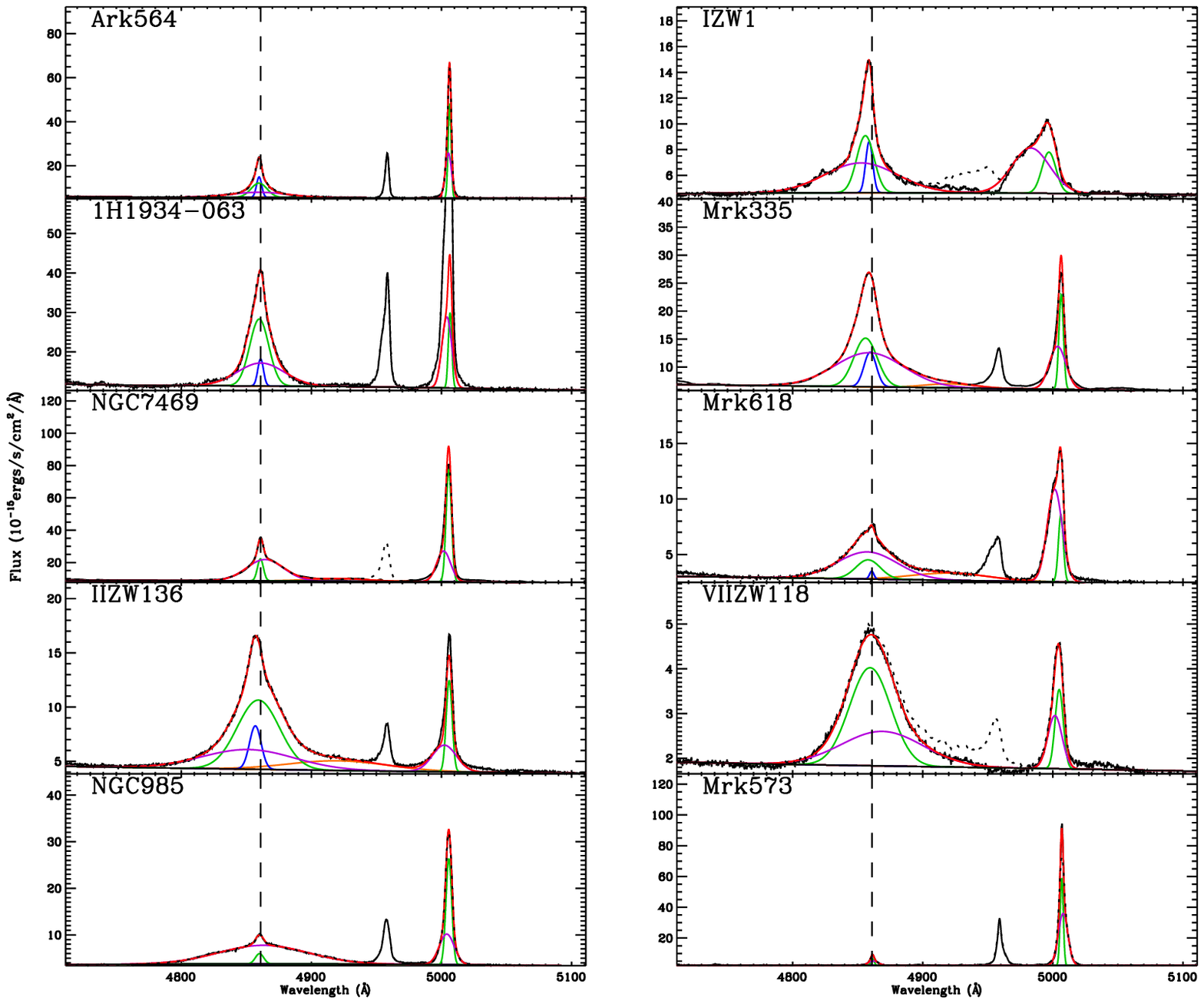}
\end{center}
\end{figure*}

\begin{figure*}
\caption{He I$\lambda$5876- IIZW136 appears to show evidence of a strong He I line, but unfortunately it could not be accurately measured, as its profile is interrupted by a gap introduced by the limited spectral range of each CCD.}
\label{HeI}
\begin{center}
	\includegraphics[width=18cm, height =
	13cm]{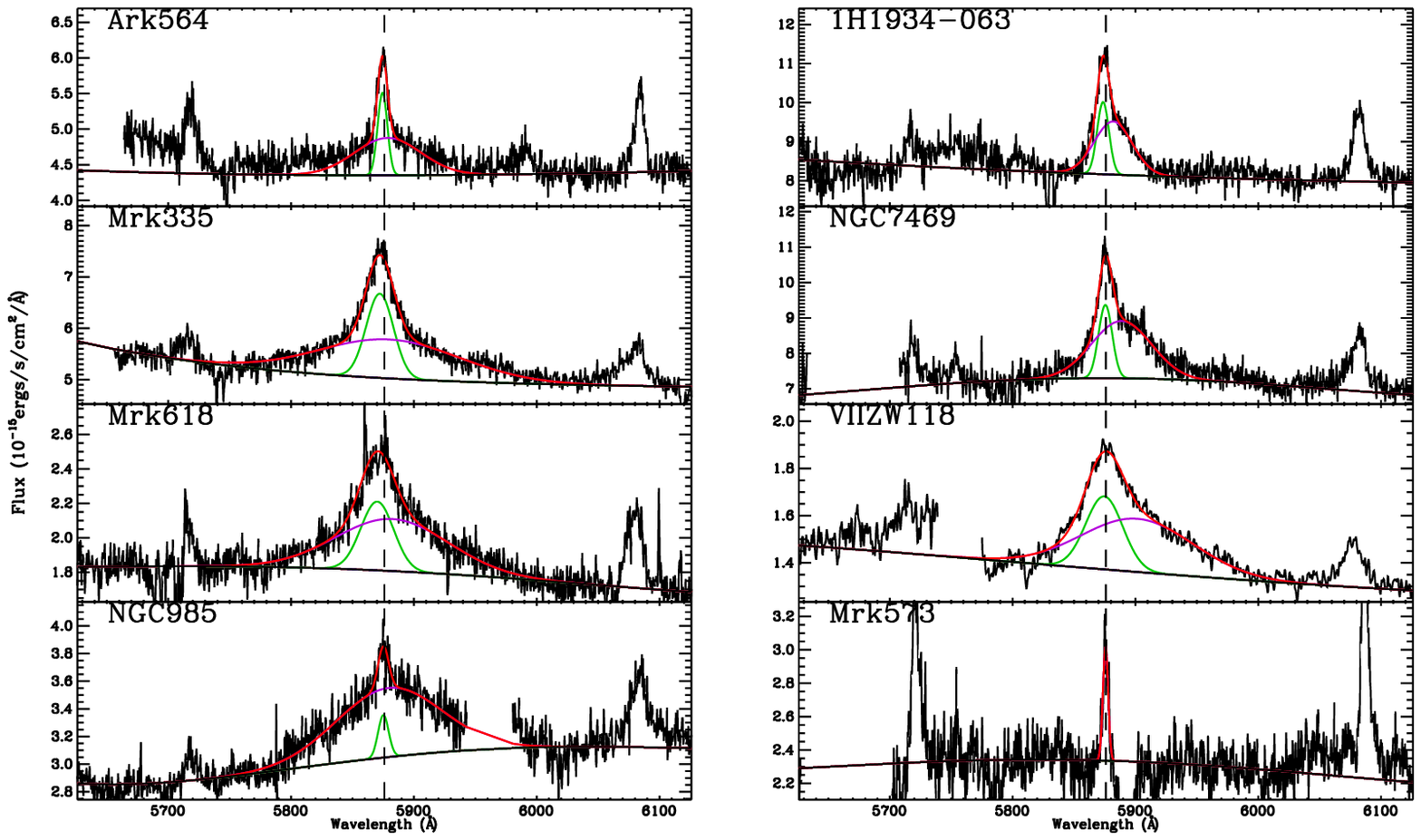}
\end{center}
\end{figure*}

\begin{figure*}
\caption{He II$\lambda$4686- We have deblended this line from the unknown line at $\sim$4670\AA, which is clearly a separate feature in a number of our targets (e.g. IIZW136, VIIZW118)}
\label{HeII}
\begin{center}
	\includegraphics[width=18cm, height =
	15cm]{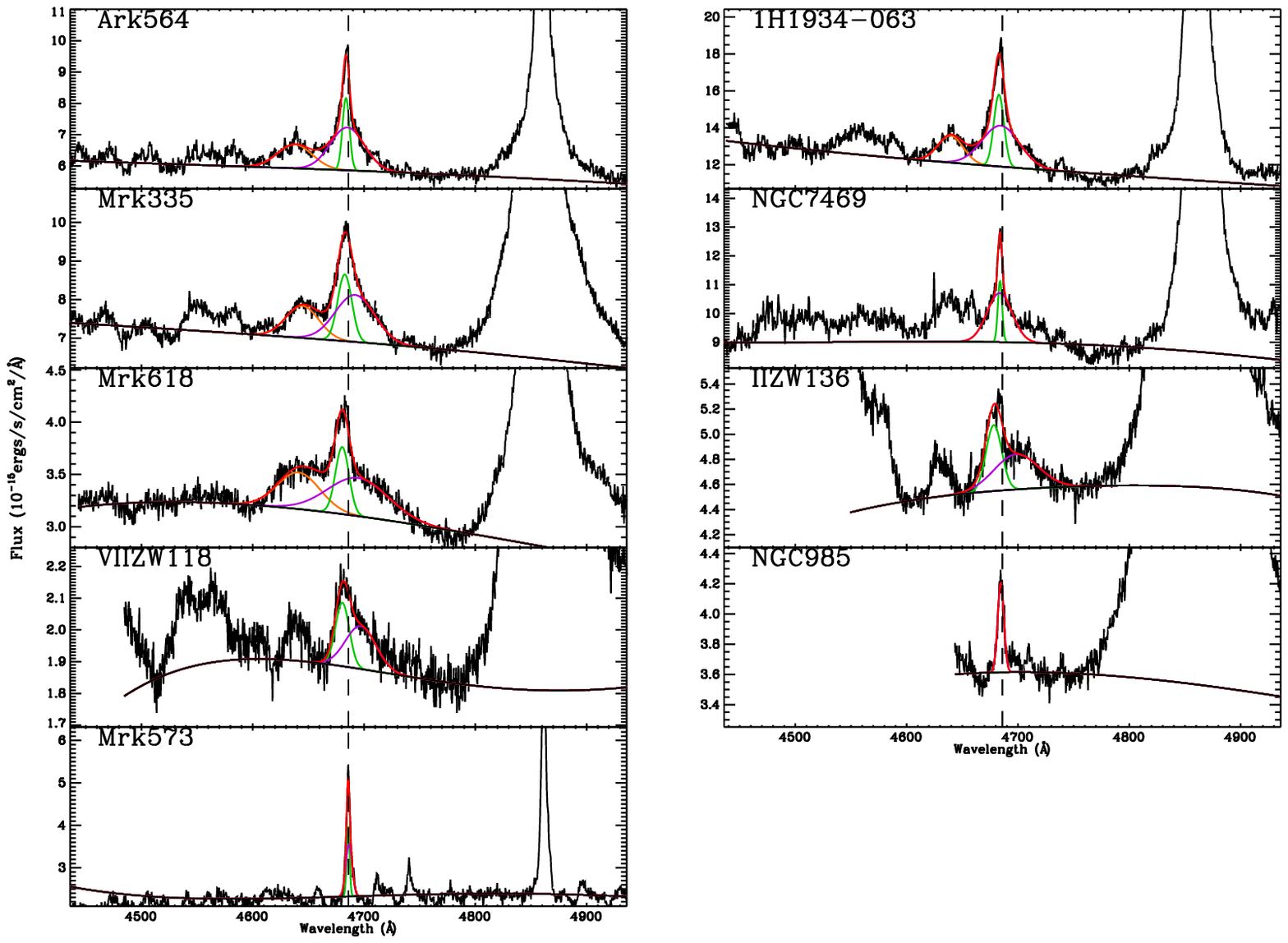}
\end{center}
\end{figure*}

\begin{figure*}
\caption{$[$Fe VII$]$$\lambda$6087}
\label{FeVII}
\begin{center}
	\includegraphics[width=18cm, height =15cm]{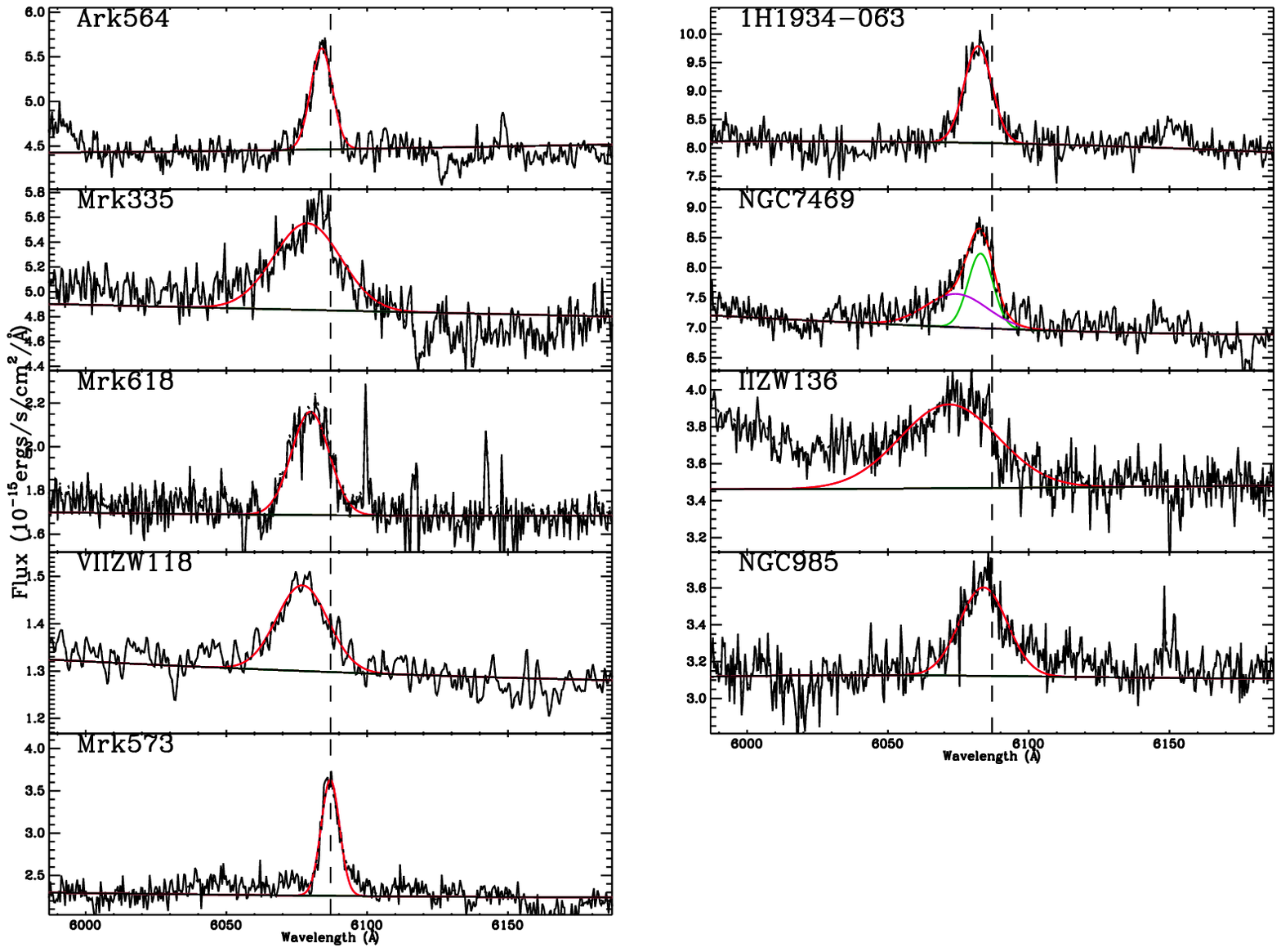}
\end{center}
\end{figure*}

\begin{figure*}
\caption{$[$Fe X$]$$\lambda$6374- Dotted lines here are used to show where we have removed both the [O I] line that is blended with the [Fe X] line.  Redward of some of the [Fe X] lines we have removed the blue wing of the H$\alpha$ line that contaminated the [Fe X] line in some of the 'broader' NLS1s in our sample.}
\label{FeX}
\begin{center}
	\includegraphics[width=18cm, height =13cm]{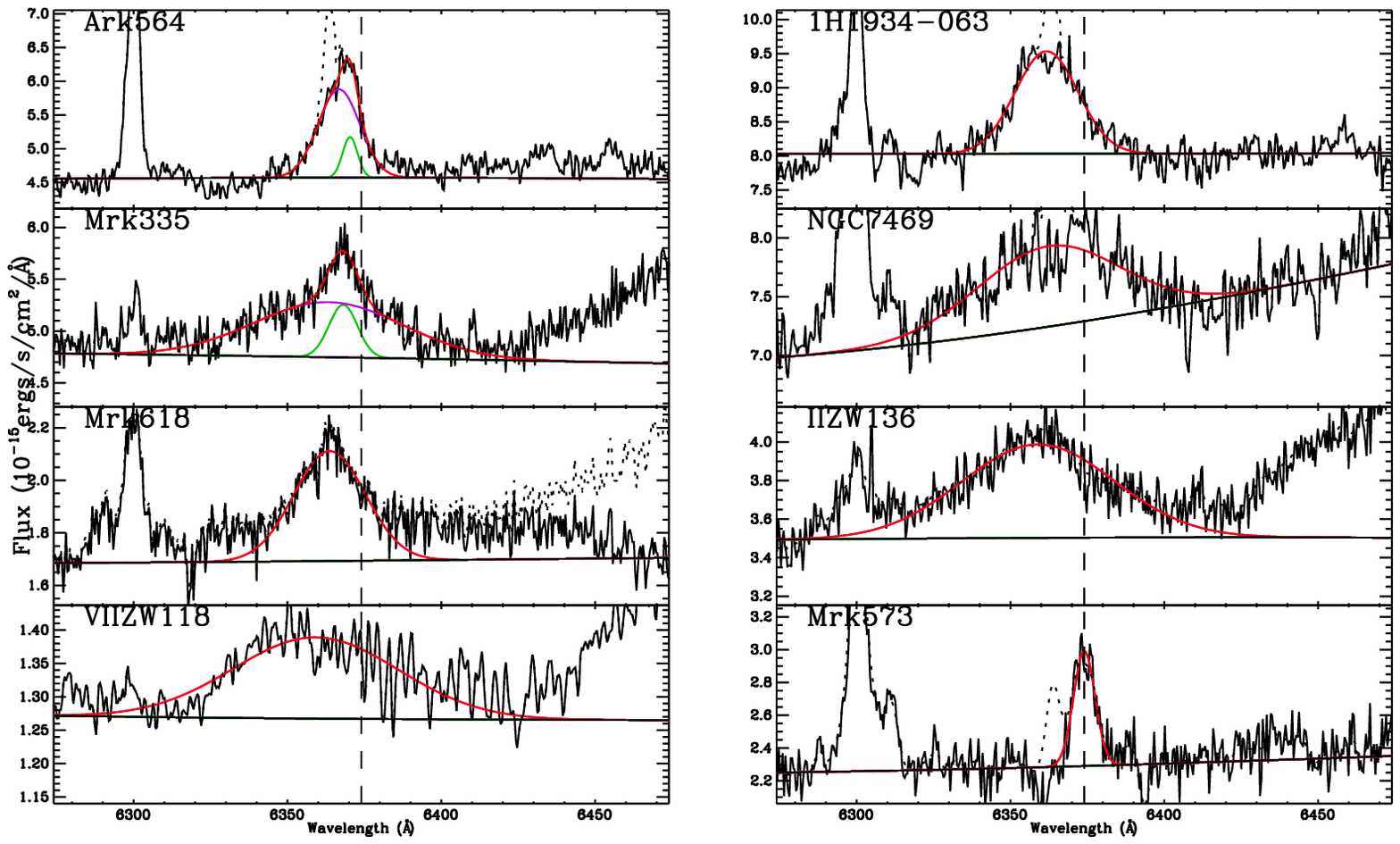}
\end{center}
\end{figure*}

\begin{figure*}
\caption{$[$Fe XI$]$$\lambda$7892}
\label{FeXI}
\begin{center}
	\includegraphics[width=18cm, height =
	7cm]{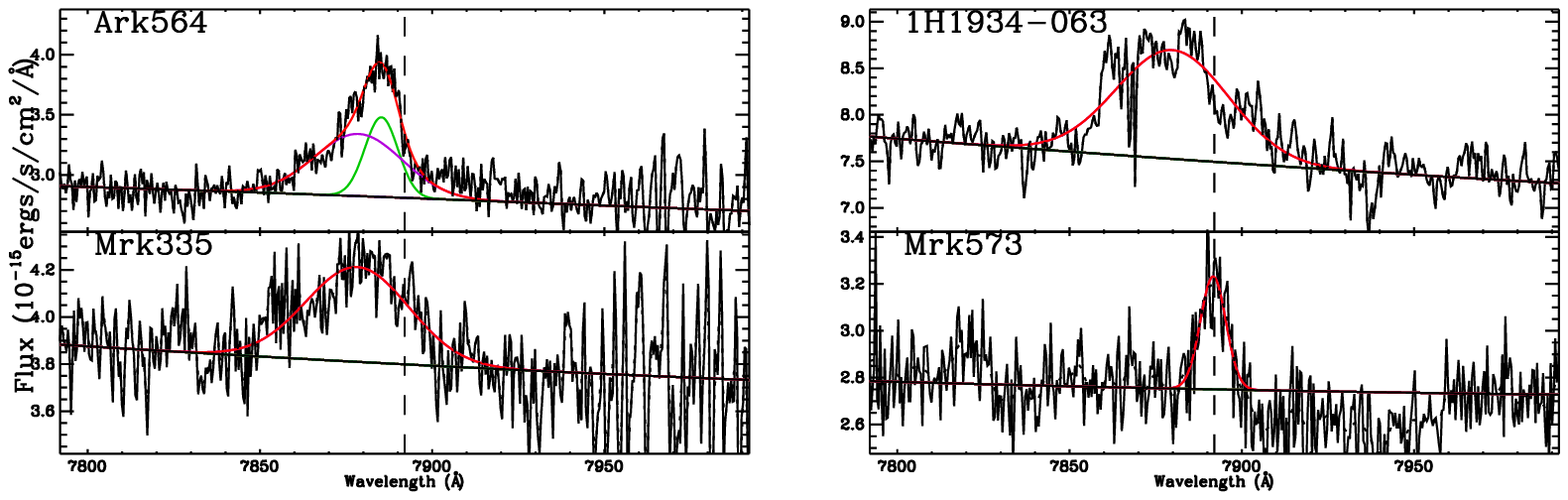}
\end{center}
\end{figure*}

\begin{figure*}
\caption{$[$Fe XIV$]$$\lambda$5303-   For Ark564, we have attempted to deblend this line from the [Ca V] line at 5309\AA\ (shown here in orange).}
\label{FeXIV}
\begin{center}
	\includegraphics[width=18cm, height =
	4cm]{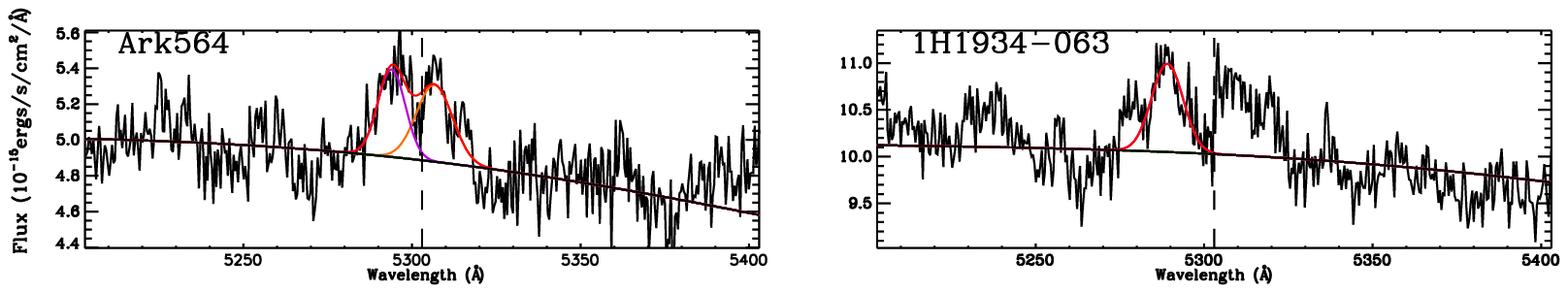}
\end{center}
\end{figure*}

\label{lastpage}

\end{document}